# Backyard Cuckoo Hashing:
# Constant Worst-Case Operations with a Succinct Representation

Yuriy Arbitman*    Moni Naor†    Gil Segev‡

September 8, 2018

**Abstract**

The performance of a dynamic dictionary is measured mainly by its update time, lookup time, and space consumption. In terms of update time and lookup time there are known constructions that guarantee constant-time operations in the worst case with high probability, and in terms of space consumption there are known constructions that use essentially optimal space. However, although the first analysis of a dynamic dictionary dates back more than 45 years ago (when Knuth analyzed linear probing in 1963), the trade-off between these aspects of performance is still not completely understood. In this paper we settle two fundamental open problems:

- We construct the first dynamic dictionary that enjoys the best of both worlds: it stores $n$ elements using $(1 + \epsilon)n$ memory words, and guarantees constant-time operations in the worst case with high probability. Specifically, for any $\epsilon = \Omega((\log \log n/ \log n)^{1/2})$ and for any sequence of polynomially many operations, with high probability over the randomness of the initialization phase, all operations are performed in constant time which is independent of $\epsilon$.

  The construction is a two-level variant of cuckoo hashing, augmented with a "backyard" that handles a large fraction of the elements, together with a de-amortized perfect hashing scheme for eliminating the dependency on $\epsilon$.

- We present a variant of the above construction that uses only $(1 + o(1))\mathcal{B}$ bits, where $\mathcal{B}$ is the information-theoretic lower bound for representing a set of size $n$ taken from a universe of size $u$, and guarantees constant-time operations in the worst case with high probability, as before. This problem was open even in the *amortized* setting. One of the main ingredients of our construction is a permutation-based variant of cuckoo hashing, which significantly improves the space consumption of cuckoo hashing when dealing with a rather small universe.

*Department of Computer Science and Applied Mathematics, Weizmann Institute of Science, Rehovot 76100, Israel. Email: `yuriy.arbitman@gmail.com`.

†Incumbent of the Judith Kleeman Professorial Chair, Department of Computer Science and Applied Mathematics, Weizmann Institute of Science, Rehovot 76100, Israel. Email: `moni.naor@weizmann.ac.il`. Research supported in part by a grant from the Israel Science Foundation. Part of this work was done while visiting the Center for Computational Intractability at Princeton University.

‡Department of Computer Science and Applied Mathematics, Weizmann Institute of Science, Rehovot 76100, Israel. Email: `gil.segev@weizmann.ac.il`. Research supported by the Adams Fellowship Program of the Israel Academy of Sciences and Humanities.

# 1 Introduction

A dynamic dictionary is a data structure used for maintaining a set of elements under insertions, deletions, and lookup queries. The first analysis of a dynamic dictionary dates back more than 45 years ago, when Knuth analyzed linear probing in 1963 [Knu63] (see also [Knu98]). Over the years dynamic dictionaries have played a fundamental role in computer science, and a significant amount of research has been devoted for their construction and analysis.

The performance of a dynamic dictionary is measured mainly by its update time, lookup time, and space consumption. Although each of these performance aspects alone can be made essentially optimal rather easily, it seems to be a highly challenging task to construct dynamic dictionaries that enjoy good performance in all three aspects. Specifically, in terms of update time and lookup time there are known constructions that guarantee constant-time operations in the worst case with high probability[1] (e.g., [DMadH90, DDM+05, DMadHP+06, ANS09]), and in terms of space consumption there are known constructions that provide almost full memory utilization (e.g., [FPS+05, Pan05, DW07]) – even with constant-time lookups, but without constant-time updates.

In this paper we address the task of constructing a dynamic dictionary that enjoys optimal guarantees in all of the above aspects. This problem is motivated not only by the natural theoretical insight that its solution may shed on the feasibility and efficiency of dynamic dictionaries, but also by practical considerations. First, the space consumption of dictionary is clearly a crucial measure for its applicability in the real world. Second, whereas amortized performance guarantees are suitable for a very wide range of applications, for other applications it is highly desirable that all operations are performed in constant time in the worst case. For example, in the setting of hardware routers and IP lookups, routers must keep up with line speeds and memory accesses are at a premium [BM01, KM07]. An additional motivation for the construction of dictionaries with worst case guarantees is combatting "timing attacks", first suggested by Lipton and Naughton [LN93]. They showed that timing information may reveal sensitive information on the randomness used by the data structure, and this can enable an adversary to identify elements whose insertion results in poor running time. The concern regarding timing information is even more acute in a cryptographic environment with an active adversary who might use timing information to compromise the security of the system (see, for example, [Koc96, TOS10]).

## 1.1 Our Contributions

In this paper we settle two fundamental open problems in the design and analysis of dynamic dictionaries. We consider the unit cost RAM model in which the elements are taken from a universe of size $u$, and each element can be stored in a single word of length $w = \lceil \log u \rceil$ bits. Any operation in the standard instruction set can be executed in constant time on $w$-bit operands. This includes addition, subtraction, bitwise Boolean operations, left and right bit shifts by an arbitrary number of positions, and multiplication[2]. Our contributions are as follows:

**Achieving the best of both worlds.** We construct a two-level variant of cuckoo hashing [PR04] that uses $(1 + \epsilon)n$ memory words, where $n$ is the maximal number of elements stored at any point in time, and guarantees constant-time operations in the worst case with high probability.

---

[1] More specifically, for any sequence of operations, with high probability over the randomness of the initialization phase of the data structure, each operation is performed in constant time.

[2] The unit cost RAM model has been the subject of much research, and is considered the standard model for analyzing the efficiency of data structures (see, for example, [DP08, Hag98, HMP01, Mil99, PP08, RR03] and the references therein).

Specifically, for any $0 < \epsilon < 1$ and for any sequence of polynomially many operations, with overwhelming probability over the randomness of the initialization phase, all insertions are performed in time $O(\log(1/\epsilon)/\epsilon^2)$ in the worst case. Deletions and lookups are always performed in time $O(\log(1/\epsilon)/\epsilon^2)$ in the worst case.

We then show that this construction can be augmented with a de-amortized perfect hashing scheme, resulting in a dynamic dictionary in which all operations are performed in constant time which is independent of $\epsilon$, for any $\epsilon = \Omega((\log \log n / \log n)^{1/2})$. The augmentation is based on a rather general de-amortization technique that can rely on any perfect hashing scheme with two natural properties.

**Succinct representation.** The above construction stores $n$ elements using $(1 + o(1))n$ memory words, which are $(1 + o(1))n \log u$ bits. This may be rather far from the information-theoretic bound of $\mathcal{B}(u, n) = \lceil \log \binom{u}{n} \rceil$ bits for representing a set of size $n$ taken from a universe of size $u$. We present a variant of our construction that uses only $(1+o(1))\mathcal{B}$ bits[3], and guarantees constant-time operations in the worst case with high probability as before. Our approach is based on hashing elements using permutations instead of functions. We first present a scheme assuming the availability of truly random permutations, and then show that this assumption can be eliminated by using $k$-wise $\delta$-dependent permutations.

**Permutation-based cuckoo hashing.** One of the main ingredients of our construction is a permutation-based variant of cuckoo hashing. This variant improves the space consumption of cuckoo hashing by storing $n$ elements using $(2+\epsilon)n \log(u/n)$ bits instead of $(2+\epsilon)n \log u$ bits. When dealing with a rather small universe, this improvement to the space consumption of cuckoo hashing might be much more significant than that guaranteed by other variants of cuckoo hashing that store $n$ elements using $(1 + \epsilon)n \log u$ bits [FPS+05, Pan05, DW07]. Analyzing our permutation-based variant is more challenging than analyzing the standard cuckoo hashing, as permutations induce inherent dependencies among the outputs of different inputs (these dependencies are especially significant when dealing with a rather small universe). Our analysis relies on subtle coupling argument between a random function and a random permutation, that is enabled by a specific monotonicity property of the bipartite graphs underlying the structure of cuckoo hashing.

**Application of small universes: A nearly-optimal Bloom filter alternative.** The difference between using $(1+o(1)) \log \binom{u}{n}$ bits and using $(1 + o(1))n \log u$ bits is significant when dealing with a small universe. An example for an application where the universe size is small and in which our construction yields a significant improvement arises when applying dictionaries to solve the *approximate set membership problem*: representing a set of size $n$ in order to support lookup queries, allowing a false positive rate of at most $0 < \delta < 1$, and no false negatives. In particular, we are interested in the *dynamic* setting where the elements of the set are specified one by one via a sequence of insertions. This setting corresponds to applications such as graph exploration where the inserted elements correspond to nodes that have already been visited (e.g. [CVW+92]), global deduplication-based compression systems where the inserted elements correspond to data segments that have already been compressed (e.g. [ZLP08]), and more. In these applications $\delta$ has to be roughly $1/n$ so as not to make any error in the whole process.

The information-theoretic lower bound for the space required by any solution to this problem is $n \log(1/\delta)$ bits, and this holds even in the static setting where the set is given in advance [CFG+78]. The problem was first solved using a Bloom filter [Blo70], whose space consumption is $n \log(1/\delta) \log e$ bits (i.e., this is a compact representation). See more in Appendix A.

---

[3]Demaine [Dem07] classifies data structures into "implicit" (redundancy $O(1)$), "succinct" (redundancy $o(\mathcal{B})$) and "compact" (redundancy $O(\mathcal{B})$).

Using our succinctly-represented dictionary we present the first solution to this problem whose space consumption is only $(1 + o(1))n \log(1/\delta) + O(n + \log u)$ bits, and guarantees constant-time lookups and insertions in the worst case with high probability (previously such guarantees were only known in the amortized sense). In particular, the lookup time and insertion time are independent of $\delta$. For any sub-constant $\delta$ (the case in the above applications), and under the reasonable assumption that $u \leq 2^{O(n)}$, the space consumption of our solution is $(1 + o(1))n \log(1/\delta)$, which is optimal up to an additive lower order term (i.e., this is a succinct representation)[4].

## 1.2 Related Work

A significant amount of work was devoted to constructing dynamic dictionaries over the years, and here we focus only on the results that are most relevant to our setting.

**Dynamic dictionaries with constant-time operations in the worst case.** Dietzfelbinger and Meyer auf der Heide [DMadH90] constructed the first dynamic dictionary with constant-time operations in the worst case with high probability, and $O(n)$ memory words for storing $n$ elements (the construction is based on the dynamic dictionary of Dietzfelbinger et al. [DKM+94]). While this construction is a significant theoretical contribution, it may be unsuitable for highly demanding applications. Most notably, it suffers from large multiplicative constant factors in its memory utilization and running time, and from an inherently hierarchal structure. Recently, Arbitman et al. [ANS09] presented a de-amortization of cuckoo hashing that guarantees constant-time operations in the worst case with high probability, and achieves memory utilization of about 50%. Their experimental results indicate that the scheme is efficient, and provides a practical alternative to the construction of Dietzfelbinger and Meyer auf der Heide.

**Dynamic dictionaries with full memory utilization.** Linear probing is the most classical hashing scheme that offers full memory utilization. When storing $n$ elements using $(1+\epsilon)n$ memory words, its expected insertion time is polynomial in $1/\epsilon$. However, for memory utilization close to 100% it is rather inefficient, and the average search time becomes linear in the number of elements stored (for more details we refer the reader to Theorem K and the subsequent discussion in [Knu98, Chapter 6.4]).

Cuckoo hashing [PR04] achieves memory utilization of slightly less than 50%, and its generalizations [FPS+05, Pan05, DW07] were shown to achieve full memory utilization. These generalizations follow two lines: using multiple hash functions, and storing more than one element in each bin. To store $n$ elements using $(1 + \epsilon)n$ memory words, the expected insertion time when using multiple hash functions was shown to be $(1/\epsilon)^{O(\log \log(1/\epsilon))}$, and when using bins with more than one element it was shown to be $\log(1/\epsilon)^{O(\log \log(1/\epsilon))}$. For further and improved analysis see also [CSW07, DM09, DGM+10, FR07, FP09, FM09, FMM09, LP09].

Fotakis et al. [FPS+05] suggested a general approach for improving the memory utilization of a given scheme by employing a multi-level construction: their dictionary comprises of several levels of decreasing sizes, and elements that cannot be accommodated in any of these levels are placed in an auxiliary dictionary. Their scheme, however, does not efficiently support deletions, and the number of levels (and thus also the insertion time and lookup time) depends on the overall loss in memory utilization.

---

[4]For constant $\delta$ there is a recent lower bound of Lovett and Porat [LP10] showing we cannot get to $(1 + o(1))n \log(1/\delta)$ bits in the dynamic setting.

**Dictionaries approaching the information-theoretic space bound.** A number of dictionaries with space consumption that approaches the information-theoretic space bound are known. Raman and Rao [RR03] constructed a dynamic dictionary that uses $(1 + o(1))\mathcal{B}$ bits, but provides only amortized guarantees and does not support deletions efficiently. The above mentioned construction of Dietzfelbinger and Meyer auf der Heide [DMadH90] was extended by Demaine et al. [DMadHP$^+$06] to a dynamic dictionary that uses $O(\mathcal{B})$ bits[5], where each operation is performed in constant time in the worst case with high probability. Of particular interest to our setting is their construction of quotient hash functions, that are used to hash elements similarly to the way our construction uses permutations (permutations can be viewed as a particular case of quotient hash functions). Our approach using $k$-wise almost independent permutations can be used to significantly simplify their construction, and in addition it allows a more uniform treatment without separately considering different ranges of the parameters.

In the static dictionary case (with no insertions or deletions) much work was done on succinct data structures. The first to achieve a succinct representation of static dictionary supporting $O(1)$ retrievals were Brodnik and Munro [BM99]. More efficient schemes were given by [Pag01] and [DP08]. Most recently, Pătraşcu [Păt08] showed a succinct dictionary where the redundancy can be $O(n/\text{polylog}(n))$.

### 1.3 Paper Organization

The remainder of this paper is organized as follows. In Section 2 we briefly overview several tools that are used in our constructions. In Section 3 we present our first construction and analyze its performance. In Section 4 we augment it with a de-amortized perfect hashing to eliminate the dependency on $\epsilon$. In Section 5 we present our second construction, which is a variant of our first construction, whose memory consumption matches the information-theoretic space bound, up to additive lower order terms. In Section 6 we present several concluding remarks and open problems. In Appendix A we propose an alternative to Bloom filters that is based on our constructions, and in Appendix B we discuss the notion of negatively related random variables which is used as a tool in our analysis.

## 2 Preliminaries and Tools

**$k$-wise independent functions.** A collection $\mathcal{F}$ of functions $f : U \to V$ is $k$-wise independent if for any distinct $x_1, \ldots, x_k \in U$ and for any $y_1, \ldots, y_k \in V$ it holds that

$$\Pr[f(x_1) = y_1 \wedge \cdots \wedge f(x_k) = y_k] = 1/|V|^k \ .$$

More generally, a collection $\mathcal{F}$ is $k$-wise $\delta$-dependent if for any distinct $x_1, \ldots, x_k \in U$ the distribution $(f(x_1), \ldots, f(x_k))$ where $f$ is sampled from $\mathcal{F}$ is $\delta$-close in statistical distance to the distribution $(f^*(x_1), \ldots, f^*(x_k))$ where $f^*$ is a truly random function. A simple example for $k$-wise independent functions is the collection of all polynomials of degree $k - 1$ over a finite field.

In this paper we are interested in functions that have a short representation and can be evaluated in constant time in the unit cost RAM model. Although there are no such constructions of $k$-wise independent functions, Siegel [Sie04] constructed a pretty good approximation that is sufficient for our applications (see also the recent improvement of Dietzfelbinger and Rink [DR09] to Siegel's construction). For any two sets $U$ and $V$ of size polynomial in $n$, and for any constant $c > 0$, Siegel presented a randomized algorithm outputting a collection $\mathcal{F}$ of functions $f : U \to V$ with the following guarantees:

---

[5]Using the terminology of Demaine [Dem07], this data structure is "compact".

1. With probability at least $1 - n^{-c}$, the collection $\mathcal{F}$ is $n^\alpha$-wise independent for some constant $0 < \alpha < 1$ that depends on $|U|$ and $n$.

2. Any function $f \in \mathcal{F}$ is represented using $n^\beta$ bits, for some constant $\alpha < \beta < 1$, and evaluated in constant time in the unit cost RAM model.

Several comments are in place regarding the applicability of Siegel's construction in our setting. First, whenever we use $n^\alpha$-wise independent functions in this paper, we instantiate them with Siegel's construction, and this contributes at most an additive $n^{-c}$ factor to the failure probability of our schemes[6]. Second, the condition that $U$ and $V$ are of polynomial size does not hurt the generality of our results: in our applications $|V| \leq |U|$, and $U$ can always be assumed to be of sufficiently large polynomial size by using a pairwise (almost) independent function mapping $U$ to a set of polynomial size without any collisions with high probability. Finally, each function is represented using $n^\beta$ bits, for some constant $\beta < 1$, and this enables us in particular to store any constant number of such functions: the additional space consumption is only $O(n^\beta) = o(n \log(u/n))$ bits which is negligible compared to the space consumption of our schemes.

A significantly simpler and more efficient construction, but with a weaker guarantee on the randomness, was provided by Dietzfelbinger and Woelfel [DW03] following Pagh and Pagh [PP08] (see also [DR09]). For any two sets $U$ and $V$ of size polynomial in $n$, and for any integer $k \leq n$ and constant $c > 0$, they presented a randomized algorithm outputting a collection $\mathcal{F}$ of functions $f : U \to V$ with the following guarantees:

1. For any *specific* set $S \subset U$ of size $k$, there is an $n^{-c}$ probability of failure (i.e., choosing a "bad" function for this set), but if failure does not occur, then a randomly chosen $f \in \mathcal{F}$ is fully random on $S$.

2. Any function $f \in \mathcal{F}$ is represented using $O(k \log n)$ bits, and evaluated in constant time in the unit cost RAM model.

Note that such a guarantee is indeed slightly weaker than that provided by Siegel's construction: in general, we cannot identify a bad event whose probability is polynomially small in $n$, so that if it does not occur then the resulting distribution is $k$-wise independent. Therefore it is harder to plug in such a distribution instead of an exact $k$-wise independent distribution (e.g., it is not clear that the $k$-th moments remain the same). Specifically, this type of guarantee implies that for a set of size $n$, if one considers all its subsets of size $k$, then a randomly chosen function from the collection behaves close to a truly random function on each set, but this does not necessary hold *simultaneously* for all subsets of size $k$, as we would like in many applications. Nevertheless, inspired by the approach of [DR09], in Section 5.4 we show that our constructions can in fact rely on such a weaker guarantee, resulting in significantly simpler and more efficient instantiations.

**$k$-wise almost independent permutations.** A collection $\Pi$ of permutations $\pi : U \to U$ is $k$-wise $\delta$-dependent if for any distinct $x_1, \ldots, x_k \in U$ the distribution $(\pi(x_1), \ldots, \pi(x_k))$ where $\pi$ is sampled from $\Pi$ is $\delta$-close in statistical distance to the distribution $(\pi^*(x_1), \ldots, \pi^*(x_k))$ where $\pi^*$ is a truly random permutation. For $k > 3$ no explicit construction is known for $k$-wise *exactly independent* permutations (i.e., $\delta = 0$), and therefore it seems rather necessary to currently settle for *almost independence* (see [KNR09] for a more elaborated discussion).

In Section 5.3 we observe a construction of $k$-wise $\delta$-dependent permutations with a short description and constant evaluation time. The construction is obtained by combining known results from two independent lines of research: constructions of pseudorandom permutations (see, for example, [LR88, NR99]), and constructions of $k$-wise independent functions with short descriptions and constant evaluation time as discussed above.

---
[6]Note that property 1 above is stronger in general than $k$-wise $\delta$-dependence.

**Deviation inequalities for random variables with limited independence.** Our analysis in this paper involves bounding tail probabilities for sums of random variables. For independent random variables these are standard applications of the Chernoff-Hoeffding bounds. In some cases, however, we need to deal with sums of random variables that are dependent, and in these cases we use two approaches. The first approach, due to Schmidt et al. [SSS95, Theorem 5], is using tail bounds for the sum of $n$ random variables that are $k$-wise independent for some $k < n$. Schmidt et al. proved that for an appropriate choice of $k$ it is possible to recover the known bounds. The second approach, due to Janson [Jan93], is to prove that the random variables under consideration are *negatively related*. Informally, this means that if some of the variables obtain higher values than expected, then the other variables obtain lower values than expected. We elaborate more on this approach in Appendix B.

## 3 The Backyard Construction

Our construction is based on two-level hashing, where the first level consists of a collection of bins of constant size each, and the second level consists of cuckoo hashing. One of the main observations underlying our construction is that the specific structure of cuckoo hashing enables a very efficient interplay between the two levels.

**Full memory utilization via two-level hashing.** Given an upper bound $n$ on the number of elements stored at any point in time, and a memory utilization parameter $0 < \epsilon < 1$, set $d = \lceil c \log(1/\epsilon)/\epsilon^2 \rceil$ for some constant $c > 1$, $m = \lceil (1 + \epsilon/2)n/d \rceil$, and $k = \lceil n^\alpha \rceil$ for some constant $0 < \alpha < 1$. The first level of our dictionary is a table $T_0$ containing $m$ entries (referred to as bins), each of which contains $d$ memory words. The table is equipped with a hash function $h_0 : \mathcal{U} \to [m]$ that is sampled from a collection of $k$-wise independent hash functions (see Section 2 for constructions of such functions with succinct representations and constant evaluation time). Any element $x \in \mathcal{U}$ is stored either in the bin $T_0[h_0(x)]$ or in the second level. The lookup procedure is straightforward: when given an element $x$, perform a lookup in the bin $T_0[h_0(x)]$ and in the second level. The deletion procedure simply deletes $x$ from its current location. As for inserting an element $x$, if the bin $T_0[h_0(x)]$ contains less than $d$ elements then we store $x$ there, and otherwise we store $x$ in the second level. We show that the number of elements that cannot be stored in the first level after exactly $n$ insertions is at most $\epsilon n/16$ with high probability. Thus, the second level should be constructed to store only $\epsilon n/16$ elements.

**Supporting deletions efficiently: cuckoo hashing.** When dealing with long sequences of operations (as opposed to only $n$ insertions as considered in the previous paragraph), we must be able to move elements from the second level back to the first level. Otherwise, when elements are deleted from the first level, and new elements are inserted into the second level, it is no longer true that the second level contains at most $\epsilon n/16$ elements at any point in time. One possible solution to this problem is to equip each first-level bin with a doubly-linked list, pointing to all the "overflowing" elements of the bin (these elements are stored in the second level). Upon every deletion from a bin in the first level we move one of these overflowing elements from the second level to this bin. We prefer, however, to avoid such a solution due to its extensive usage of pointers and the rather inefficient maintenance of the linked lists.

We provide an efficient solution to this problem by using cuckoo hashing as the second level dictionary. Cuckoo hashing uses two tables $T_1$ and $T_2$, each consisting of $r = (1 + \delta)\ell$ entries for some small constant $\delta > 0$ for storing at most $\ell = \epsilon n/16$ elements, and two hash functions $h_1, h_2 : \mathcal{U} \to \{1, \ldots, r\}$. An element $x$ is stored either in entry $h_1(x)$ of table $T_1$ or in entry $h_2(x)$ of table $T_2$, but never in both. The lookup and deletion procedure are naturally defined, and as for insertions, Pagh and Rodler [PR04] proved that the "cuckoo approach", kicking other elements away until every element has its own "nest", leads to an efficient insertion procedure. More specifically, in order to insert an element $x$ we store it in entry $T_1[h_1(x)]$. If this entry is not occupied, then we are done, and otherwise we make its previous occupant "nestless". This element is then inserted to $T_2$ using $h_2$ in the same manner, and so forth iteratively. We refer the reader to [PR04] for a more comprehensive description of cuckoo hashing.

A very useful property of cuckoo hashing in our setting is that in its insertion procedure, whenever stored elements are encountered we add a test to check whether they actually "belong" to the main table $T_0$ (i.e., whether their corresponding bin has an available entry). The key property is that if we ever encounter such an element, the insertion procedure is over (since an available position is found for storing the current nestless element). Therefore, as far as the cuckoo hashing is concerned, it stores at most $\epsilon n/16$ elements at any point in time. This guarantees that any insert operation leads to at most one insert operation in the cuckoo hashing, and one insert operation in the first-level bins.

**Constant worst-case operations: de-amortized cuckoo hashing.** Instead of using the classical cuckoo hashing we use the recent construction of Arbitman et al. [ANS09] who showed how to de-amortize the insertion time of cuckoo hashing using a queue. The insertion procedure in the second level is now parameterized by a constant $L$, and is defined as follows. Given a new element $x$ (which cannot be stored in the first level), we place the pair $(x, 1)$ at the *back* of the queue (the additional value indicates to which of the two cuckoo tables the element should be inserted next). Then, we carry out the following procedure as long as no more than $L$ moves are performed in the cuckoo tables: we take the pair $(y, b)$ from the *head* of the queue, and check whether $y$ can be inserted into the first level. If its bin in the first level is not full then we store $y$ there, and otherwise we place $y$ in entry $T_b[h_b(y)]$. If this entry was unoccupied (or if $y$ was successfully moved to the first level of the dictionary), then we are done with the current element $y$, this is counted as one move and the next element is fetched from the *head* of the queue. However, if the entry $T_b[h_b(y)]$ was occupied, we check whether its previous occupant $z$ can be stored in the first level and otherwise we store $z$ in entry $T_{3-b}[h_{3-b}(z)]$ and so on, as in the above description of the standard cuckoo hashing. After $L$ elements have been moved, we place the current "nestless" element at the *head* of the queue, together with a bit indicating the next table to which it should be inserted, and terminate the insertion procedure (note that it may take less than $L$ moves, if the queue becomes empty). An important ingredient in the construction of Arbitman et al. is the implicit use of a small auxiliary data structure called "stash" that enables to avoid rehashing, as suggested by Kirsch et al. [KMW09].

A schematic diagram of our construction is presented in Figure 1, and a formal description of its procedures is provided in Figure 2.

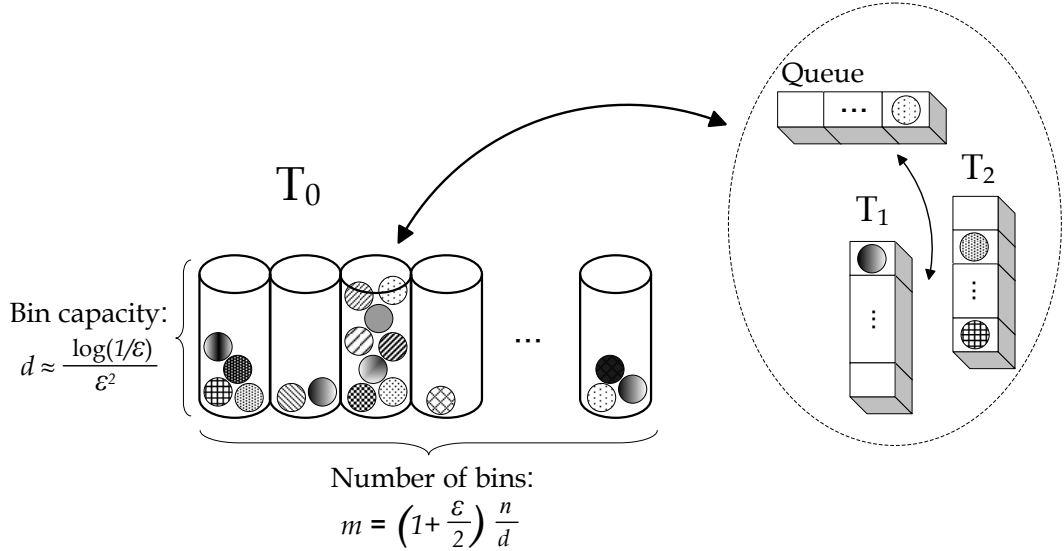

**Figure 1:** A schematic diagram of the backyard scheme.

We prove the following theorem:

**Theorem 3.1.** *For any $n$ and $0 < \epsilon < 1$ there exists a dynamic dictionary with the following properties:*

1. *The dictionary stores $n$ elements using $(1+\epsilon)n$ memory words.*

2. *For any polynomial $p(n)$ and for any sequence of at most $p(n)$ operations in which at any point in time at most $n$ elements are stored in the dictionary, with probability at least $1 - 1/p(n)$ over the randomness of the initialization phase, all insertions are performed in time $O(\log(1/\epsilon)/\epsilon^2)$ in the worst case. Deletions and lookups are always performed in time $O(\log(1/\epsilon)/\epsilon^2)$ in the worst case.*

**Proof.** We first compute the total number of memory words used by our construction. The main table $T_0$ consists of $m = \lceil (1+\epsilon/2)n/d \rceil$ entries, each of which contains $d$ memory words. The de-amortized cuckoo hashing is constructed to store at most $\epsilon n/16$ elements at any point in time. Arbitman et al. [ANS09] showed that the de-amortized cuckoo hashing achieves memory utilization of $1/2 - \delta$ for any constant $0 < \delta < 1$, and for our purposes it suffices to assume, for example, that it uses $\epsilon n/4$ memory words. Thus, the total number of memory words is $md + \epsilon n/4 \leq (1+\epsilon)n$.

In the remainder of the proof we analyze the correctness and performance of our construction. We show that it suffices to construct the de-amortized cuckoo hashing under the assumption that it does not contain more than $\epsilon n/16$ elements at any point in time, and that we obtain the worst-case performance guarantees stated in the theorem. The technical ingredient in this argument is a lemma stating a bound on the number of "overflowing" elements when $n$ elements are placed in $m$ bins using a $k$-wise independent hash function $h : \mathcal{U} \to [m]$. Specifically, we say that an element is overflowing if it is mapped to a bin together with at least $d$ other elements.

We follow essentially the same analysis presented in [PP08, Section 4]. For the following lemma recall that $d = \lceil c \log(1/\epsilon)/\epsilon^2 \rceil$, $m = \lceil (1+\epsilon/2)n/d \rceil$, and $k = \lceil n^\alpha \rceil$ for some constant $0 < \alpha < 1$. Given a set $S \subset \mathcal{U}$ we denote by $Q = Q(S) \subset S$ the set of elements that are placed in bins with no more than $d-1$ other elements (i.e., $Q$ is the set of non-overflowing elements).

```
Lookup(x):                                          Delete(x):
 1: if x is stored in bin h_0(x) of T_0 then         1: if x is stored in bin h_0(x) of T_0 then
 2:     return true                                  2:     Remove x from bin h_0(x)
 3: else                                             3: else
 4:     return LookupCuckoo(x)                       4:     DeleteFromCuckoo(x)

Insert(x):
 1: InsertIntoBackOfQueue(x, 1)
 2: y ←⊥ // y denotes the current element
 3: for i = 1 to L do // L denotes the number of permitted moves in cuckoo tables
 4:     if y =⊥ then // Fetching element y from the head of the queue
 5:         if IsQueueEmpty() then
 6:             return
 7:         else
 8:             (y, b) ← PopFromQueue()
 9:     if there is vacant place in bin h_0(y) of T_0 then
10:         Store y in bin h_0(y) of T_0
11:         y ←⊥
12:     else
13:         if T_b[h_b(y)] =⊥ then // Successful insert
14:             T_b[h_b(y)] ← y
15:             ResetCDM()
16:             y ←⊥
17:         else
18:             if LookupInCDM(y, b) then // Found the second cycle
19:                 InsertIntoBackOfQueue(y, b)
20:                 ResetCDM()
21:                 y ←⊥
22:             else // Evict existing element
23:                 z ← T_b[h_b(y)]
24:                 T_b[h_b(y)] ← y
25:                 InsertIntoCDM(y, b)
26:                 y ← z
27:                 b ← 3 − b
28: if y ≠⊥ then
29:     InsertIntoHeadOfQueue(y, b)
```

**Figure 2:** The procedures of the backyard scheme.

**Lemma 3.2.** *For any set $S \subset \mathcal{U}$ of size $n$, with probability $1 - 2^{-\omega(\log n)}$ over the choice of a $k$-wise independent hash function $h : \mathcal{U} \to [m]$, it holds that $|Q(S)| \geq (1 - \epsilon/16)n$.*

**Proof.** Denote by $B_i \subseteq S$ the set of elements that are placed in the $i$-th bin, and denote by $z = \omega(\log n)$ the largest integer for which $2^z d \leq k$. Split the set of bins $[m]$ into blocks of at most $2^z$ consecutive bins $I_j = \{2^z j + 1, \ldots, 2^z(j+1)\}$, for $j = 0, 1, \ldots, m/2^z - 1$. Without loss of generality we assume that $2^z$ divides $m$. Otherwise the last block contains less than $2^z$ bins, and we can count all the elements that are mapped to these bins as overflowing, and as the remainder of the proof shows this will have only a negligible effect on the size of the set $Q$.

First, we argue that for every block $I_j$ it holds that $|\cup_{i \in I_j} B_i| \leq (1-\epsilon/4)2^z d$ with probability $1 - 2^{-\omega(\log n)}$. We prove this by using a Chernoff bound for random variables with limited independence due to Schmidt et al. [SSS95]. Fix some block $I_j$, and for any element $x \in S$ denote by $Y_{x,j}$ the indicator random variable of the event in which $x$ is placed in one of the bins of block $I_j$, and let $Y_j = \sum_{x \in S} Y_{x,j}$. Each indicator $Y_{x,j}$ has expectation $2^z/m$, and thus $E(Y_j) = 2^z n/m$. In addition, these indicators are $k$-wise independent. Therefore,

$$
\begin{aligned}
\Pr\left[|\cup_{i \in I_j} B_i| > \left(1 - \frac{\epsilon}{4}\right) 2^z d\right] &= \Pr\left[Y_j > \left(1 - \frac{\epsilon}{4}\right) 2^z d\right] \\
&\leq \Pr\left[Y_j > \left(1 - \frac{\epsilon}{4}\right)\left(1 + \frac{\epsilon}{2}\right) E(Y_j)\right] \\
&\leq \Pr\left[Y_j > \left(1 + \frac{\epsilon}{8}\right) E(Y_j)\right] \\
&\leq \Pr\left[|Y_j - E(Y_j)| > \frac{\epsilon}{8} \cdot E(Y_j)\right] \\
&\leq \exp\left(-\left(\frac{\epsilon}{8}\right)^2 \cdot \frac{E(Y_j)}{3}\right) \\
&= 2^{-\omega(\log n)} ,
\end{aligned}
\tag{3.1}
$$

where (3.1) follows from [SSS95, Theorem 5.I.b] by our choice of $k$.

Now, assuming that for every block $I_j$ it holds that $|\cup_{i \in I_j} B_i| \leq (1 - \epsilon/4) 2^z d$, we argue that for every block $I_j$ it holds that $|Q \cap (\cup_{i \in I_j} B_i)| \geq (1 - \epsilon/16)|\cup_{i \in I_j} B_i|$ with probability $1 - 2^{-\omega(\log n)}$, and this concludes the proof of the lemma.

Fix the value of $j$, and note that our choice of $z$ such that $2^z d \leq k$ implies that the values of $h$ on the elements mapped to block $I_j$ are completely independent. Therefore we can apply a Chernoff bound for completely independent random variables to obtain that for any $i \in I_j$ it holds that

$$\Pr\left[||B_i| - E(|B_i|)| > \frac{\epsilon}{32} E(|B_i|)\right] \leq 2e^{-\Omega(\epsilon^2 d)} .$$

Denote by $\overline{Z}_i$ the indicator random variable of the event in which $|B_i| > (1 + \epsilon/32) E(|B_i|)$, and by $\underline{Z}_i$ the indicator random variable of the event in which $|B_i| < (1 - \epsilon/32) E(|B_i|)$. Although the random variables $\{\overline{Z}_i\}_{i \in I_j}$ are not independent, they are *negatively related* (see Appendix B for more details), and this allows us to apply a Chernoff bound on their sum. The same holds for the random variables $\{\underline{Z}_i\}_{i \in I_j}$, and therefore we obtain that with probability $1 - 2^{-\omega(\log n)}$ for at least $(1 - 4e^{-\Omega(\epsilon^2 d)}) 2^z$ bins from the block $I_j$ it holds that $(1 - \epsilon/32)|\cup_{i \in I_j} B_i|/2^z \leq |B_i| \leq d$. The elements stored in these bins are non-overflowing, and therefore the number of non-overflowing elements in this block is at least $(1 - \epsilon/32)(1 - 4e^{-\Omega(\epsilon^2 d)})|\cup_{i \in I_j} B_i|$. The choice of $d = O(\log(1/\epsilon)/\epsilon^2)$ implies that the latter is at least $(1 - \epsilon/16)|\cup_{i \in I_j} B_i|$. ∎

Consider now a sequence of at most $p(n)$ operations such that at any point in time the dictionary contains at most $n$ elements. For every $1 \leq i \leq p(n)$ denote by $S_i$ the set of elements that are stored in the dictionary after the execution of the first $i$ operations, and denote by $A_i \subseteq S_i$ the set of elements that are mapped by the function $h_0$ of the first-level table to bins that contain more than $d$ elements from the set $S_i$ (i.e., using the terminology of Lemma 3.2, $A_i$ is the set of overflowing elements when the elements in the set $S_i$ are placed in the first level). Then, Lemma 3.2 guarantees that for every $1 \leq i \leq p(n)$ it holds that $|A_i| \leq \epsilon n/16$ with probability $1 - 2^{-\omega(\log n)}$. A union bound then implies that again with probability $1 - 2^{-\omega(\log n)}$ it holds that $|A_i| \leq \epsilon n/16$ for every $1 \leq i \leq p(n)$. We now show that this suffices for obtaining the worst-case performance guarantees:

**Lemma 3.3.** *Assume that for every $1 \leq i \leq p(n)$ it holds that $|A_i| \leq \epsilon n/16$ (i.e., there are at most $\epsilon n/16$ overflowing elements at any point in time). Then with probability at least $1 - 1/p(n)$ over the randomness used in the initialization phase of the de-amortized cuckoo hashing, insertions are performed in time $O(\log(1/\epsilon)/\epsilon^2)$ in the worst case. Deletions and lookups are always performed in time $O(\log(1/\epsilon)/\epsilon^2)$ in the worst case.*

**Proof.** The insertion procedure is defined such that whenever it runs into an element in the second level that can be stored in the first level, then this element is moved to the first level, and either an available position is found in one of the second-level tables or the queue of the second level shrinks by one element. In both of these cases we turn to deal with the next element in the queue. Thus, we can compare the insertion time in the second level to that of the de-amortized cuckoo hashing: as far as our insertion procedure is concerned, the elements that are effectively stored at any point in time in the second level are a subset of $A_i$ (the set of overflowing elements after the $i$-th operation), and each operation in the original insertion procedure is now followed by an access to the first level. Any access to the first level takes time linear in the size $d = \lceil c \log(1/\epsilon)/\epsilon^2 \rceil$ of a bin, and therefore with probability $1 - 1/p(n)$ each insert operation is performed in time $O(\log(1/\epsilon)/\epsilon^2)$ in the worst case. As for deletions and lookups, they are always performed in time linear in the size $d = \lceil c \log(1/\epsilon)/\epsilon^2 \rceil$ of a bin. ■

This concludes the proof of Theorem 3.1. ■

## 4 De-amortized Perfect Hashing: Eliminating the Dependency on $\epsilon$

The dependency on $\epsilon$ in the deletion and lookup times can be eliminated by using a perfect hashing scheme (with a succinct representation) in each of the first-level bins. Upon storing an element in one of the bins, the insertion procedure reconstructs the perfect hash function for this bin. As long as the reconstruction can be done in time linear in the size of a bin, then the insertion procedure still takes time $O(d) = O(\log(1/\epsilon)/\epsilon^2)$ in the worst case, and the deletion and lookup procedures take constant time that is independent of $\epsilon$. Such a solution, however, does not eliminate the dependency on $\epsilon$ in the insertion time.

In this section we present an augmentation that completely eliminates the dependency on $\epsilon$. We present a rather general technique for de-amortizing a perfect hashing scheme to be used in each of the first-level bins. Our approach relies on the fact that the same scheme is employed in a rather large number of bins at the same time, and this enables us to use a queue to guarantee that even insertions are performed in constant time that is independent of $\epsilon$. Using this augmentation we immediately obtain the following refined variant of Theorem 3.1 (the restriction $\epsilon = \Theta((\log \log n / \log n)^{1/2})$ is due to the specific scheme that we de-amortize – see more details below):

**Theorem 4.1.** *For any integer $n$ there exists a dynamic dictionary with the following properties:*

1. *The dictionary stores $n$ elements using $(1+\epsilon)n$ memory words, for $\epsilon = \Theta((\log \log n / \log n)^{1/2})$.*

2. *For any polynomial $p(n)$ and for any sequence of at most $p(n)$ operations in which at any point in time at most $n$ elements are stored in the dictionary, with probability at least $1 - 1/p(n)$ over the randomness of the initialization phase, all operations are performed in constant time, independent of $\epsilon$, in the worst case.*

This augmentation is rather general and we can use any perfect hashing scheme with two natural properties. We require that for any sequence $\sigma$ of operations leading to a set $S$ of size at most $d-1$, for any sequence of memory configurations and rehashing times occurring during the execution of $\sigma$, and for any element $x \notin S$ that is currently being inserted it holds that:

**Property 1:** With probability $1 - O(1/d)$ the current hash function can be adjusted to support the set $S \cup \{x\}$ in expected constant time. In addition, the adjustment time in this case is always upper bounded by $O(d)$.

**Property 2:** With probability $O(1/d)$ rehashing is required, and the rehashing time is dominated by $O(d) \cdot Z$ where $Z$ is a geometric random variable with a constant expectation.

Our augmentation introduces an overhead which imposes a restriction on the range of possible values for $\epsilon$. The restriction comes from two sources: the description length of the perfect hash function in every bin, and the computation time of the hash function and its adjustment on every insertion. We propose a specific scheme that satisfies the above properties, and can handle $\epsilon = \Omega((\log \log n / \log n)^{1/2})$. It is rather likely that various other schemes such as [FKS84, DKM+94] can be slightly modified to satisfy these properties. In particular, the schemes [Pag99, Woe06] seem especially suitable for this purpose.

To de-amortize any scheme that satisfies these two properties we use an auxiliary queue (the same queue is used for all bins), and the insertion procedure to the bins is now defined as follows: upon insertion, the new element is always placed at the *back* of the queue, and then we perform a constant number of steps (denoted by $L$) on the element currently located at the *head* of the queue. If these $L$ steps are not enough to insert this element into its bin, we return it to the *head* of the queue, and continue working on this element upon the next insertion. If we managed to insert this element by using less than $L$ steps, we continue with the next element and so on until we complete $L$ steps[7]. As for deletions, these are also processed using the queue, and when deleting an element we simply locate the element inside its bin and mark it as deleted (i.e., deletions are always performed in constant time).

The key point in the analysis is that properties 1 and 2 guarantee that the expected amount of work for each element is a small constant, which in turn implies that the queue does not grow beyond $O(\log n)$ with high probability. Specifically, we show that the constant number of operations that we perform upon every insertion can be chosen independently of $\epsilon$ such that with high probability the queue is always of size $O(\log n)$. Thus, as long as the queue does not overflow, all operations are performed in constant time that is independent of $\epsilon$. In what follows we formally prove that with high probability the queue does not overflow.

Consider a sequence of at most $p(n)$ operations, for some polynomial $p(n)$, such that at most $n$ elements are stored in the data structure at any point in time. Fix the first-level hash function $h_0$, and denote by $\sigma = (x_1, \ldots, x_N)$ the sequence of operations on the first-level bins in reverse order (each operation is either insertion or deletion depending on whether the element is currently stored or not). For any element $x_i$ denote by $W(x_i)$ the total amount of work required for storing $x_i$ in its bin (note that elements may appear more than once).

**Lemma 4.2.** *For any constant $c_1 > 0$ and any integer $T$ there exists a constant $c_2$, such that for any $1 \leq i_0 \leq N - T$ it holds that*

$$\Pr\left[\sum_{i=1}^{T} W(x_{i_0+i}) \geq c_2 T\right] \leq \exp(-c_1 T/d) \ .$$

---

[7]A comment is in place regarding rehashing. If rehashing is needed, then we copy the content of the rehashed bin to a dedicated memory location, perform the rehash, and then copy back the content of the bin, and all this is done in several phases of $L$ steps. Note that the usage of the queue guarantees that at any point in time we rehash at most one bin.

**Proof.** For simplicity we let $W_i = W(x_{i_0+i})$, and assume that all $T$ operations are insertions (as discussed above, deletions are always performed in constant time). We argue that although the $W_i$'s are not independent, they are nevertheless dominated by independent random variables with the same distribution. First, note that since different bins use independently chosen perfect hash functions, then given the allocation of elements into bins (i.e., conditioned on the function $h_0$ of the first level), $W_i$'s that correspond to different bins are independent. Consider now a pair $W_i$ and $W_j$ for which the elements $x_i$ and $x_j$ are mapped to the same bin, and assume without loss of generality that $x_i$ is processed from the head of the queue before $x_j$. Then by the time we process $x_j$, we either already adjusted the hash function to store $x_i$, or we are already done with the rehashing of the bin due to $x_i$ (this follows from the fact that we always return the current element we work with to the *head* of the queue). Properties 1 and 2 hold for any memory configurations and rehashing times (in particular, those possibly caused by $x_j$), and therefore $W_i$ and $W_j$ are dominated by independent random variables with the same distribution as guaranteed by these two properties (that is, the time it takes to process $x_j$ can be assumed to be independent of the time it took to process $x_i$). Note that this argument is actually not limited to considering only pairs, and thus we conclude that $W_1, \ldots, W_T$ are dominated by independent random variables with the same distribution.

We split the elements $x_1, \ldots, x_T$ into two sets: those that cause rehashing, and those that does not cause rehashing. Property 2 implies that the expected number of elements that cause rehashing is at most $\alpha T/d$, for some constant $\alpha$, and thus a Chernoff bound guarantees that with probability $1 - \exp(-\Omega(T/d))$ at most $2\alpha T/d$ elements cause rehashing. For these elements a concentration bound for the sum of i.i.d. geometric random variables (also known as the negative binomial distribution[8]) with expectation $O(d)$ implies that with probability $1 - \exp(-\Omega(T/d))$ the sum of their corresponding $W_i$'s does not exceed $O(T)$.

As for the remaining elements (i.e., those that do not cause rehashing), property 1 and the above discussion guarantee that the sum of their corresponding $W_i$'s is dominated by sum of $T$ i.i.d. random variables with support $\{1, \ldots, O(d)\}$ and constant expectation. Thus, the Hoeffding bound guarantees that with probability $1 - \exp(-\Omega(T/d^2))$ their sum does not exceed $O(T)$. ∎

Denote by $\mathcal{E}$ the event in which for every $1 \leq j \leq N/\log n$ it holds that

$$\sum_{i=1}^{\log n} W(x_{(j-1)\log n+i}) \leq c_2 \log n \ .$$

An appropriate choice of the constant $c_1$ in Lemma 4.2 and a union bound imply that the event $\mathcal{E}$ occurs with probability at least $1 - n^{-c}$, for any pre-specified constant $c$. The following lemma bounds the size of the queue assuming that the event $\mathcal{E}$ occurs.

**Claim 4.3.** *Assuming that the event $\mathcal{E}$ occurs, then during the execution of $\sigma$ the queue does not contain more than $2\log n$ elements at any point in time.*

**Proof.** We prove by induction on $j$, that at the time $x_{j\log n+1}$ is inserted into the queue, there are no more than $\log n$ elements in the queue. This clearly implies that at any point in time there are at most $2\log n$ elements in the queue. For $j = 1$ we observe that at most $\log n$ elements were inserted into the first level. In particular, there can be at most $\log n$ elements in the queue.

Assume that the statement holds for some $j$, and we prove that it holds also for $j + 1$. The inductive hypothesis states that at the time $x_{j\log n+1}$ is inserted, the queue contains at most $\log n$ elements. In the worst case, these elements are $\{x_{(j-1)\log n+1}, \ldots, x_{j\log n}\}$ (it is possible that the element at the head of the queue is replaced by another element from its bin due to rehashing, but this only means that a certain amount of work was already devoted for that operation). Therefore, the event $\mathcal{E}$ implies that the elements $\{x_{(j-1)\log n+1}, \ldots, x_{j\log n}\}$ can be handled in $c_2 \log n$ steps. By choosing the constant $L$ such that $L \log n \geq c_2 \log n$ (recall that $L$ is the number of steps that we complete on each operation), it is guaranteed that by the time the element $x_{(j+1)\log n+1}$ is inserted into the queue, these $\log n$ elements are already processed. Thus, by the time the element $x_{(j+1)\log n+1}$ is inserted into the queue, the queue contains at most the elements $\{x_{j\log n+1}, \ldots, x_{(j+1)\log n}\}$ (where, again, the element at the head of the queue may be replaced by another element from its bin due to rehashing). ∎

---

[8]See, for example, [DP09, Problem 2.4].

Finally, we note that there are several possibilities for implementing the queue with constant time deletions and lookups. Here we adopt the suggestion of Arbitman et al. [ANS09] and use a constant number $\ell$ of arrays $A_1, \ldots, A_\ell$ each of size $n^\delta$, for some $\delta < 1$. Each entry of these arrays consists of a data element, a pointer to the previous element in the queue, and a pointer to the next element in the queue. The elements are stored using a function $\widehat{h}$ chosen from a collection of pairwise independent hash functions. We refer the reader to [ANS09] for more details.

## 4.1 A Specific Scheme for $\epsilon = \Omega((\log \log n / \log n)^{1/2})$

The scheme uses exactly $d$ memory words to store $d$ elements, and 3 additional words to store the description of its hash function. The elements are mapped into the set $[d]$ using two functions. The first is a pairwise independent function $h$ mapping the elements into the set $[d^2]$. This function can be described using 2 memory words and evaluated in constant time. The second is a function $g$ that records for each $r \in [d^2]$ for which there is a stored element $x$ with $h(x) = r$ the location of $x$ in $[d]$. The description of $g$ consists of at most $d$ pairs taken from $[d^2] \times [d]$ and therefore can be represented using $3d \log d$ bits.

The lookup operation of an element $x$ computes $h(x) = r$ and then $g(r)$ to check if $x$ is stored in that location. In general, we cannot assume that the function $g$ can be evaluated in constant time, and therefore we also store a lookup table for its evaluation. This table is shared by all the bins, and it represents the function that takes as input the description of $g$ and a value $r$, and outputs $g(r)$ or null. The size of this lookup table is $2^{3d \log d + 2 \log d} \cdot \log d$ bits. The deletion operation performs a lookup for $x$, and then updates the description of $g$. Again, for updating the description of $g$ we use another lookup table (shared among all bins) that takes as input the current description of $g$ and a value $r = h(x)$, and outputs a new description for $g$. The size of this lookup table is $2^{3d \log d + 2 \log d} \cdot 3d \log d$ bits.

As for the insert operation, in Claim 4.4 below we prove that with probability $1 - O(1/d)$ a new element will not introduce a collision for the function $h$. In this case we store the new element in the next available entry of $[d]$, and update the description of $g$. For identifying the next available entry we use a global lookup table of size $2^d \log d$ bits (each row in the table corresponds to an array of $d$ bits describing the occupied entries of a bin), and for updating the description of $g$ we use a lookup table of size $2^{3d \log d + 2 \log d} \cdot 3d \log d$ bits as before. With probability $O(1/d)$ when inserting a new element we need to rehash by sampling a new function $h$, and executing the insert operation on all the elements. In this case the rehashing time is upper bounded by $O(d) \cdot Z$ where $Z$ is a geometric random variable with a constant expectation. Thus, this scheme satisfies the two properties stated in the beginning of the section.

The total amount of space used by the global lookup tables is $O(2^{3d \log d + 2 \log d} \cdot d \log d)$ bits. For $\epsilon = \Omega((\log \log n / \log n)^{1/2})$ this is at most $n^\alpha$ bits for some constant $0 < \alpha < 1$, and therefore negligible compared to our space consumption. In addition, the hash function of every bin is described using $2\log u + d\log d$ bits, and therefore summing over all $m = \lceil(1+\epsilon/2)n/d\rceil$ bins this is $O(n/d \cdot \log u + n\log d)$. For $\epsilon = \Omega(\log\log n / \log n)$ this is at most $\epsilon n \log u$ bits, which is again negligible compared to our space consumption. Thus, this forces the restriction $\epsilon = \Omega((\log\log n / \log n)^{1/2})$.

For simplifying the proof of the following claim we introduce a "forced rehashing" condition into our scheme. We add to the description of the hash function in every bin an integer $\nu \in \{1, \ldots, d\}$ that is chosen uniformly at random, and we always rehash after $\nu$ update operations, unless we rehashed sooner due to a collision in the function $h$. On every rehashing we choose a new value $\nu$. Note that this increases the probability of rehashing in every update operation by an additive term of $1/d$, and this does not hurt properties 1 and 2.

**Claim 4.4.** *Let $1 \leq \ell < d$, fix a sequence $\sigma$ of operations leading to set $S$ of size $\ell$, and assume that $S$ does not have any collisions under the currently chosen function $h : \mathcal{U} \to [d^2]$. Then, for any sequences of memory configurations and rehashing times that occurred during the execution of $\sigma$, and for any element $x \notin S$, the probability over the choice of $h$ that $x$ will form a collision with an element of $S$ is $O(1/d)$.*

**Proof.** Assume first that there are only insertions and no deletions. Then the current hash function $h$ is uniformly distributed in the collection of pairwise independent functions subject to not having any collisions in the set $S$. Therefore, for any element $x \notin S$ it holds that

$$\Pr[x \text{ collides with an element of } S \mid h \text{ is 1-1 on } S]$$
$$= \frac{\Pr[x \text{ collides with an element of } S \wedge h \text{ is 1-1 on } S]}{\Pr[h \text{ is 1-1 on } S]}$$
$$\leq \frac{\Pr[x \text{ collides with an element of } S]}{\Pr[h \text{ is 1-1 on } S]} .$$

The function $h$ is chosen from a collection of pairwise independent hash functions, and therefore

$$\Pr[x \text{ collides with an element of } S] \leq |S|/d^2 \leq 1/d ,$$

and

$$\Pr[h \text{ is 1-1 on } S] \geq 1 - d(d-1)/2d^2 \geq 1/2 .$$

These implies that

$$\Pr[x \text{ collides with an element of } S \mid h \text{ is 1-1 on } S] \leq 2/d .$$

When dealing with both insertions and deletions, it is no longer true that the current hash function is uniformly distributed subject to not having any collisions in the set $S$. However, since we always rehash after at most $d$ update operations, then even if we ignore all deletions since the last rehash (i.e., we include in the set $S$ all the deleted elements since the last rehash) we are left with a set of size at most $3d/2$, for which the latter is true, and the same analysis as above holds. ∎

## 5 Matching the Information-Theoretic Space Bound

In this section we present a variant of our construction that uses only $(1 + o(1))\mathcal{B}$ bits, where $\mathcal{B} = \mathcal{B}(u, n)$ is the information-theoretic bound for representing a set of size $n$ taken from a universe of size $u$, and guarantees constant-time operations in the worst case with high probability as before. We first present a scheme that is based on truly random permutations, and then present a scheme that is based on $k$-wise $\delta$-dependent permutations. Finally, we present a construction of such permutations with short descriptions and constant evaluation time. We prove the following theorem:

**Theorem 5.1.** *For any integers $u$ and $n \leq u$ there exists a dynamic dictionary with the following properties:*

1. *The dictionary stores $n$ elements taken from a universe of size $u$ using $(1+\epsilon)\mathcal{B}$ bits, where $\mathcal{B} = \lceil \log \binom{u}{n} \rceil$ and $\epsilon = \Theta(\log \log n/(\log n)^{1/3})$.*

2. *For any polynomial $p(n)$ and for any sequence of at most $p(n)$ operations in which at any point in time at most $n$ elements are stored in the dictionary, with probability at least $1 - 1/p(n)$ over the randomness of the initialization phase, all operations are performed in constant time, independent of $\epsilon$, in the worst case.*

One of the ideas we will utilize is that when we apply a permutation $\pi$ to an element $x$ we may think of $\pi(x)$ as a new identity for $x$, provided that we are also able to compute $\pi^{-1}(x)$. The advantage is that we can now store explicitly only part of $\pi(x)$, where the remainder is stored *implicitly* by the location where the value is stored. This is the idea behind quotient hash functions, as suggested previously by Pagh [Pag01] and Demaine et al. [DMadHP+06].

## 5.1 A Scheme based on Truly Random Permutations

Recall that our construction consists of two levels: a table in the first level that contains $m \approx n/d$ bins, each of which stores at most $d$ elements, and the de-amortized cuckoo hashing in the second level for dealing with the overflowing elements. The construction described in this section shares the same structure, while refining the memory consumptions in each of the two levels separately. In turn, Theorem 5.1 (assuming truly random permutations for now) follows immediately by plugging in the following modifications to our previous schemes.

### 5.1.1 First-Level Hashing Using Permutations

We reduce the space consumption in the first level of our construction by hashing the elements into the first-level table using a "chopped" permutation $\pi$ over the universe $\mathcal{U}$ as follows. For simplicity we first assume that $u$ and $m$ are powers of 2, and then we explain how to deal with the more general case. Given a permutation $\pi$ and an element $x \in \mathcal{U}$, we denote by $\pi_L(x)$ the left-most $\log m$ bits of $\pi(x)$, and by $\pi_R(x)$ the right-most $\log(u/m)$ bits of $\pi(x)$. That is, $\pi(x)$ is the concatenation of the bit-strings $\pi_L(x)$ and $\pi_R(x)$. We use $\pi_L$ as the function mapping elements into bins, and $\pi_R$ as the identity of the elements inside the bins: any element $x$ is stored either in the first level in bin $\pi_L(x)$ using the identity $\pi_R(x)$, or in the second level if its first-level bin already contains $d$ other elements. The update and lookup procedures remain exactly the same, and note that the correctness of the lookup procedure is guaranteed by the fact that $\pi$ is a permutation, and therefore the function $\pi_R$ is one-to-one inside every bin.

In the following lemma we bound the number of overflowing elements in the first level when using a truly random permutation. Recall that an element is overflowing if it is mapped to a bin with at least $d$ other elements. The lemma guarantees that by setting $d = O(\log(1/\epsilon)/\epsilon^2)$ there are at most $\epsilon n/16$ overflowing elements with an overwhelming probability, exactly as in Section 3.

**Lemma 5.2.** *Fix any $n$, $d$, $\epsilon$, and a set $S \subseteq \mathcal{U}$ of $n$ elements. With probability $1 - 2^{-\omega(\log n)}$ over the choice of a truly random permutation $\pi$, when using the function $\pi_L$ for mapping the elements* of $S$ into $m = \lceil (1+\epsilon)n/d \rceil$ *bins of size $d$, the number of* non-overflowing *elements is at least* $(1 - \epsilon/32)(1 - 4e^{-\Omega(\epsilon^2 d)})n$.

**Proof.** For any $i \in [m]$ denote by $B_i$ the number of elements that are mapped to the $i$-th bin. Each $B_i$ is distributed according to the hypergeometric distribution (i.e., random sampling *without replacement*) with expectation $n/m$, and using known concentrations results for this distribution (see, for example, [Chv79, Hoe63, SSS95]) we have that

$$\Pr\left[|B_i - E(B_i)| > \frac{\epsilon}{32} \cdot E(B_i)\right] \leq 2e^{-\Omega(\epsilon^2 d)} .$$

Denote by $\overline{I}_i$ the indicator random variable of the event in which $B_i > (1 + \epsilon/32)E(B_i)$, and by $\underline{I}_i$ the indicator random variable of the event in which $B_i < (1 - \epsilon/32)E(B_i)$. Although the random variables $\{\overline{I}_i\}_{i=1}^m$ are not independent, they are *negatively related* (see Appendix B for more details), and this allows us to apply a Chernoff bound on their sum. The same holds for the random variables $\{\underline{I}_i\}_{i=1}^m$, and therefore we obtain that with probability $1 - 2^{-\omega(\log n)}$ for at least $(1 - 4e^{-\Omega(\epsilon^2 d)})m$ bins it holds that $(1 - \epsilon/32)n/m \leq B_i \leq d$. The elements stored in these bins are non-overflowing, and therefore the number of non-overflowing elements is at least $(1 - \epsilon/32)(1 - 4e^{-\Omega(\epsilon^2 d)})n$. ∎

We now explain how to deal with the more general case in which $u$ and $m$ are not powers of 2. First, if $m$ divides $u$ then our approach naturally extends to defining $\pi_L(x) = \left\lfloor \frac{\pi(x)}{u/m} \right\rfloor$ and $\pi_R(x) = \pi(x) \mod u/m$, and the exact same analysis holds. Second, if $m$ does not divide $u$, then it seems tempting to artificially increase the universe to a universe of size $u' < u + m$ such that $m$ divides $u'$. However, when $u$ is very small compared to $n$ (specifically, when $u < 2n$), this may significantly hurt the space consumption of our construction. Therefore, instead of increasing the size of the universe, we decrease the size of the universe to $u' > u - m$ by ignoring at most $m - 1$ elements, such that $m$ divides $u'$. Then clearly $\binom{u'}{n} \leq \binom{u}{n}$, and therefore the space consumption is not hurt. However, we need to deal with the deleted elements separately if we ever encounter them. The number of such elements is less than $m$, which is significantly smaller than the number of elements in the second level, which is $\epsilon n/16$. Therefore we can simply store these elements in the second level without affecting the performance of the construction.

### 5.1.2 The Bins in the First-Level Table

We follow the general approach presented in Section 4 to guarantee that the update and lookup operations on the first-level bins are performed in constant time that is independent of the size of the bins (and thus independent of $\epsilon$). Depending on the ratio between the size of the universe $u$ and the number of elements $n$, we present hashing schemes that satisfy the two properties stated in the beginning of Section 4. Our task here is a bit more subtle than in Section 4 since we must guarantee that the descriptions of the hash functions inside the bins (and any global lookup tables that are used) do not occupy too much space compared to the information-theoretic bound. This puts a restriction on the size of the bins. We consider two cases (these cases are not necessarily mutually exclusive):

**Case 1: $u \leq n \cdot 2^{(\log n)^\beta}$ for some $\beta < 1$.** In this case we store all elements in a single word using the information-theoretic representation, and use lookup tables to guarantee constant time operations. Specifically, recall that the elements in each bin are now taken from a universe of size $u/m$, and each bin contains at most $d$ elements. Thus, the content of a bin can be represented using $\lceil \log \binom{u/m}{d} \rceil$ bits. Insertions and deletions are performed using a global lookup table that is shared among all bins. The table represents a function that receives as input a description of a bin, and an additional element, and outputs an updated description for the bin. This lookup table can be represented using $2^{\lceil \log \binom{u/m}{d} \rceil + \lceil \log \frac{u/m}{d} \rceil} \cdot \lceil \log \binom{u/m}{d} \rceil$ bits. Similarly, lookups are performed using a global table that occupies $2^{\lceil \log \binom{u/m}{d} \rceil + \lceil \log \frac{u/m}{d} \rceil}$ bits.

These force two restrictions on $d$. First, the description of a bin has to fit into one memory word, to enable constant-time evaluation using the lookup tables. Second, the two lookup tables have to fit into at most, say, $(\epsilon/6) \cdot n \log(u/n)$ bits. When assuming that $u \leq n \cdot 2^{(\log n)^\beta}$ for some $\beta < 1$, these two restrictions allow $d = O((\log n)^{1-\beta})$. Recall that $d = O(\log(1/\epsilon)/\epsilon^2)$, and this implies that $\epsilon = \Omega\left(\frac{(\log \log n)^{1/2}}{(\log n)^{(1-\beta)/2}}\right)$.

**Case 2: $u > n \cdot 2^{(\log n)^\beta}$ for some $\beta < 1$.** In this case we use the scheme described in Section 4.1. In every bin the pairwise independent function $f$ can be represented using $2\lceil \log(u/m) \rceil$ bits (as opposed to $2\lceil \log u \rceil$ bits in Section 4.1), and the function $g$ can be represented using $3d\lceil \log d \rceil$ bits (as in Section 4.1). Summing these over all $m$ bins results in $O(n/d \cdot \log(u/n) + n \log d)$ bits, and therefore the first restriction is that the latter is at most, say, $(\epsilon/12) \cdot n \log(u/n)$ bits. Assuming that $u > n \cdot 2^{(\log n)^\beta}$ for some $\beta < 1$ (and recall that $d = O(\log(1/\epsilon)/\epsilon^2)$) this allows $\epsilon = \Omega\left(\frac{\log \log n}{(\log n)^\beta}\right)$.

In addition, as discussed in Section 4.1, the scheme requires global lookup tables that occupy a total $O(2^{3d \log d + 2 \log d} \cdot d \log d)$ bits, and therefore the second restriction is that the latter is again at most $(\epsilon/12) \cdot n \log(u/n)$ bits. This allows $d = O(\log n / \log \log n)$, and therefore $\epsilon = \Omega\left(\left(\frac{\log \log n}{\log n}\right)^{1/2}\right)$. Thus, in this case we can deal with $\epsilon = \Omega\left(\max\left\{\frac{\log \log n}{(\log n)^\beta}, \left(\frac{\log \log n}{\log n}\right)^{1/2}\right\}\right)$.

An essentially optimal trade off (asymptotically) between these two cases occurs for $\beta = 1/3$, which allows $\epsilon = \Omega\left(\frac{(\log \log n)^{1/2}}{(\log n)^{1/3}}\right)$ in the first case, and $\epsilon = \Omega\left(\frac{\log \log n}{(\log n)^{1/3}}\right)$ in the second case. Therefore, regardless of the ratio between $u$ and $n$, our construction can always allow $\epsilon = \Omega\left(\frac{\log \log n}{(\log n)^{1/3}}\right)$.

### 5.1.3 The Second Level: Permutation-based Cuckoo Hashing

First of all note that if $u > n^{1+\alpha}$ for some constant $\alpha < 1$, then $\log u \leq (1/\alpha + 1) \log(u/n)$, and therefore we can allow ourselves to store $\alpha \epsilon n$ overflowing elements using $\log u$ bits each as before. For the general case, we present a variant of the de-amortized cuckoo hashing scheme that is based on permutations, where each element is stored using roughly $\log(u/n)$ bits instead of $\log u$ bits[9].

Recall that cuckoo hashing uses two tables $T_1$ and $T_2$, each consisting of $r = (1 + \delta)\ell$ entries for some small constant $\delta > 0$ for storing a set $S \subseteq \mathcal{U}$ of at most $\ell$ elements, and two hash functions $h_1, h_2 : \mathcal{U} \to [r]$. An element $x$ is stored either in entry $h_1(x)$ of table $T_1$ or in entry $h_2(x)$ of table $T_2$. This naturally defines the cuckoo graph, which is the bipartite graph defined on $[r] \times [r]$ with edges $\{(h_1(x), h_2(x))\}$ for every $x \in S$.

We modify cuckoo hashing to use permutations as follows (for simplicity we assume that $u$ and $r$ are powers of 2, but this is not essential[10]). Given two permutations $\pi_1$ and $\pi_2$ over $\mathcal{U}$, we define

---

[9]There is also an auxiliary data structure (a queue) that contains roughly $\log n$ elements, each of which can be represented using $\log u$ bits.

[10]More generally, as discussed in Section 5.1.1, it suffices that $r$ divides $u$. The choice of $r$ is flexible since the space consumption of the second level should be $O(\epsilon n)$. With our choice of $m$ and $r$, we can increase $r$ to $r' < r + m$ such that $m$ will divide $r'$, and then decrease $u$ to $u' > u - r'$ such that $r'$ will divide $u'$ (effectively ignoring at most $O(\epsilon n)$ elements that are placed in the second level if ever encountered, as suggested in Section 5.1.1).

$h_1$ as the left-most $\log r$ bits of $\pi_1$, and $h_2$ as the left-most $\log r$ bits of $\pi_2$. An element $x$ is stored either in entry $h_1(x)$ of table $T_1$ using the right-most $\log(u/r)$ bits of $\pi_1(x)$ as its new identity, or in entry $h_2(x)$ of table $T_2$ using the right-most $\log(u/r)$ bits of $\pi_2(x)$ as its new identity. The update and lookup procedures are naturally defined as before. Note that the permutations $\pi_1$ and $\pi_2$ have to be easily invertible to allow moving elements between the two tables, and this is satisfied by our constructions of $k$-wise $\delta$-dependent permutations in Section 5.3. We now argue that by slightly increasing the size $r$ of each table, the de-amortization of cuckoo hashing (and, in particular, cuckoo hashing itself) still has the same performance guarantees when using permutations instead of functions. The de-amortization of [ANS09] relies on two properties of the cuckoo graph:

1. With high probability the sum of sizes of any $\log \ell$ connected components is $O(\log \ell)$.

2. The probability that there are at least $s$ edges that close a second cycle is $O(r^{-s})$.

These properties are known to be satisfied when $h_1$ and $h_2$ are truly random functions, and here we present a coupling argument showing that they are satisfied also when $h_1$ and $h_2$ are defined as above using truly random permutations. Our argument relies on the monotonicity of these properties: if they are satisfied by a graph, then they are also satisfied by all its subgraphs. We prove the following claim:

**Claim 5.3.** *Let $\ell = \lceil \epsilon n / 16 \rceil$ and $r = \lceil (1+\delta)(1+\epsilon)\ell \rceil$ for some constant $0 < \delta < 1$. There exists a joint distribution $\mathcal{D} = (\mathcal{G}_{f_1,f_2}, \mathcal{G}_{\pi_1,\pi_2})$ such that:*

- $\mathcal{G}_{f_1,f_2}$ *is identical to the distribution of cuckoo graphs over $[r] \times [r]$ with $\lceil (1+\epsilon)\ell \rceil$ edges, defined by $h_1$ and $h_2$ that are the left-most $\log r$ bits of two truly random functions $f_1, f_2 : \mathcal{U} \to \mathcal{U}$.*

- $\mathcal{G}_{\pi_1,\pi_2}$ *is identical to the distribution of cuckoo graphs over $[r] \times [r]$ with $\ell$ edges, defined by $h_1$ and $h_2$ that are the left-most $\log r$ bits of two truly random permutations $\pi_1, \pi_2 : \mathcal{U} \to \mathcal{U}$.*

- *With probability $1 - e^{-\Omega(\epsilon^3 n)}$ over the choice of $(G_{f_1,f_2}, G_{\pi_1,\pi_2}) \leftarrow \mathcal{D}$, it holds that $G_{\pi_1,\pi_2}$ is a subgraph of $G_{f_1,f_2}$.*

**Proof.** Let $S \subseteq \mathcal{U}$ be a set containing $\ell$ elements. We describe an iterative process for adding $\ell' = \lceil (1+\epsilon)\ell \rceil$ edges one by one to the cuckoo graph on $[r] \times [r]$ defined by truly random functions $f_1, f_2 : \mathcal{U} \to \mathcal{U}$ (this specifies the distribution $\mathcal{G}_{f_1,f_2}$). During this process we identify the edges that correspond to the subgraph defined by truly random permutations $\pi_1, \pi_2 : \mathcal{U} \to \mathcal{U}$ (this specifies the distribution $\mathcal{G}_{\pi_1,\pi_2}$).

The process consists of several phases, where at the beginning the values of $f_1$, $f_2$, $\pi_1$, and $\pi_2$ are completely undefined. In the first phase we go over all the elements of $S$ (say, in lexicographical order), and for each element $x \in S$ we sample the two values $f_1(x), f_2(x) \in \mathcal{U}$ uniformly at random and independently of all previous choices. If the value $f_1(x)$ does not collide with any previously defined value $\pi_1(x')$, and the value $f_2(x)$ does not collide with any previously defined value $\pi_2(x')$, then we define $\pi_1(x) = f_1(x)$ and $\pi_2(x) = f_2(x)$. In addition, we add the edge $(h_1(x), h_2(x))$ to the cuckoo graph, where $h_1$ and $h_2$ are the left-most $\log r$ bits of $f_1$ and $f_2$, respectively. If there is a collision in at least one of $f_1(x)$ and $f_2(x)$, then we still add the edge $(h_1(x), h_2(x))$, but do not define the values $\pi_1(x)$ and $\pi_2(x)$, and $x$ is moved to the second phase, and so on. If we have completed the process of defining the values of $\pi_1$ and $\pi_2$ on $S$ by adding only $t \leq \ell'$ edges to the graph, then we add $\ell' - t$ edges uniformly at random and halt. Otherwise, if we have already added $\ell'$ edges and did not complete the process of defining the values of $\pi_1$ and $\pi_2$ on $S$, then we define $\pi_1$ and $\pi_2$ uniformly at random (as permutations) on the remaining elements.

It is straightforward that the resulting $f_1$ and $f_2$ are truly random functions, and the resulting $\pi_1$ and $\pi_2$ are truly random permutations. Moreover, as long as we completed the process of defining the values of $\pi_1$ and $\pi_2$ on $S$ by adding at most $\ell'$ edges, then the graph defined by $\pi_1$ and $\pi_2$ is contained in the graph defined by $f_1$ and $f_2$. Thus, it only remains to prove that with high probability at most $\ell'$ edges are required for defining $\pi_1$ and $\pi_2$ on $S$.

Observe that the number of edges needed for defining $\pi_1$ and $\pi_2$ on $S$ is dominated by the sum of $\ell$ i.i.d. geometric random variables with expectation $1 + \epsilon/2$. Indeed, for every element $x \in S$ denote by $Z_x$ the random variable corresponding to the number of edges that are sampled until successfully defining $\pi_1(x)$ and $\pi_2(x)$. At any point in time the permutations $\pi_1$ and $\pi_2$ are defined on at most $\ell$ elements, and therefore a union bound implies that the probability of collision with a previously defined value is at most $2\ell/u \leq \epsilon/16$, and this holds *independently* of all the other samples. Therefore, the expectation of each $Z_x$ is at most $1/(1 - \epsilon/16) \leq 1 + \epsilon/2$, and we can treat these random variables as completely independent. Therefore, a concentration bound for the sum of $\ell$ i.i.d. geometric random variables (also known as the negative binomial distribution[11]) with expectation $1 + \epsilon/2$ implies that with probability $1 - e^{-\Omega(\epsilon^2 \ell)} = 1 - e^{-\Omega(\epsilon^3 n)}$ their sum does not exceed $\ell' = \lceil(1+\epsilon)\ell\rceil$. ∎

### 5.1.4 The Total Memory Utilization

We now compute the total number of occupied bits by considering the different parts of our construction. In the first level there are $m = \lceil(1+\epsilon)n/d\rceil$ bins each storing at most $d = O(\log(1/\epsilon)/\epsilon^2)$ elements. The representation of the bins depends on the ratio between $u$ and $n$ as considered above. In both cases we showed that the overhead of describing the hash functions of the bins and the global lookup tables is at most $\epsilon \mathcal{B}/6$ bits. We now consider the two cases separately:

**Case 1: $u \leq n \cdot 2^{(\log n)^\beta}$ for some $\beta < 1$.** In this case the elements inside each bin are represented using $\lceil \log \binom{u/m}{d} \rceil$ bits, and therefore the total number of bits occupied by the elements in all $m$ bins is

$$m \cdot \left\lceil \log \binom{u/m}{d} \right\rceil \leq m \cdot \left(\log \binom{u/m}{d} + 1\right)$$

$$\leq \log \binom{u}{md} + m \qquad (5.1)$$

$$\leq \log \binom{u}{(1+\epsilon)n} + \frac{2n}{d}$$

$$\leq \log \left(\binom{u}{n} \left(\frac{u}{n}\right)^{\epsilon n}\right) + \epsilon^2 n \qquad (5.2)$$

$$= \log \binom{u}{n} + \epsilon \log \left(\frac{u}{n}\right)^n + \epsilon^2 n$$

$$\leq \log \binom{u}{n} + \epsilon \log \binom{u}{n} + \epsilon^2 n$$

$$\leq (1 + \epsilon + \epsilon^2) \mathcal{B} ,$$

where Equation (5.1) follows from the inequality $\binom{n_1}{k_1}\binom{n_2}{k_2} \leq \binom{n_1+n_2}{k_1+k_2}$, and Equation (5.2) follows from the fact that

$$\frac{\binom{u}{(1+\epsilon)n}}{\binom{u}{n}} = \frac{(u-n)\cdots(u-(1+\epsilon)n+1)}{((1+\epsilon)n)\cdots(n+1)} \leq \left(\frac{u}{n}\right)^{\epsilon n} .$$

---
[11]See, for example, [DP09, Problem 2.4].

**Case 2: $u > n \cdot 2^{(\log n)^\beta}$ for some $\beta < 1$.** In this case the elements inside each bin are represented using $d \log(u/m)$ bits, and therefore the total number of bits occupied by the elements in all $m$ bins is

$$\begin{aligned}
m \cdot d \cdot \log \frac{u}{m} &\leq \left( \frac{(1+\epsilon)n}{d} + 1 \right) \cdot d \cdot \left( \log \frac{u}{n} + \log d \right) \\
&= (1+\epsilon) n \log \frac{u}{n} + \left\{ (1+\epsilon) n \log d + d \log \frac{u}{n} + d \log d \right\} \\
&\leq (1+\epsilon) n \log \frac{u}{n} + \epsilon \cdot n \log \frac{u}{n} \\
&\leq (1+2\epsilon)\mathcal{B} ,
\end{aligned} \quad (5.3)$$

where Equation (5.3) follows from the restriction $\epsilon = \Omega(\log \log n / (\log n)^\beta)$ that is assumed in this case.

Finally, the second level uses at most $\epsilon \mathcal{B}$ bits, and therefore the total number of bits used by our construction is at most $(1+3\epsilon)\mathcal{B}$.

## 5.2 A Scheme based on $k$-wise $\delta$-dependent Permutations

We eliminate the need for truly random permutations by first reducing the problem of dealing with $n$ elements to several instances of the problem on $n^\alpha$ elements, for some $\alpha < 1$. Then, for each such instance we apply the solution that assumes truly random permutations, but using a $k$-wise $\delta$-dependent permutation, for $k = n^\alpha$ and $\delta = 1/\text{poly}(n)$, that can be shared among all instances. Although the following discussion can be framed in terms of any small constant $\alpha < 1$, for concreteness we use $\alpha \approx 1/10$.

Specifically, we hash the elements into $m = n^{9/10}$ bins of size at most $d = n^{1/10} + n^{3/40}$ each, using a permutation $\pi : \mathcal{U} \rightarrow \mathcal{U}$ sampled from a collection $\Pi$ of one-round Feistel permutations, and prove that with overwhelming probability there are no overflowing bins. The collection $\Pi$ is defined as follows. For simplicity we assume that $u$ and $m$ are powers of 2, and then we explain how to deal with the more general case. Let $\mathcal{F}$ be a collection of $k'$-wise independent functions $f : \{0,1\}^{\log(u/m)} \rightarrow \{0,1\}^{\log m}$, where $k' = O(n^{1/20})$, with a short representation and constant evaluation time (see Section 2). Given an input $x \in \{0,1\}^{\log u}$ we denote by $x_L$ its left-most $\log m$ bits, and by $x_R$ its right-most $\log(u/m)$ bits. For every $f \in \mathcal{F}$ we define a permutation $\pi = \pi_f \in \Pi$ by $\pi(x) = (x_L \oplus f(x_R), x_R)$. Any element $x$ is mapped to the bin $\pi_L(x) = x_L \oplus f(x_R)$, and is stored there using the identity $\pi_R(x) = x_R$. A schematic diagram is presented in Figure 3.

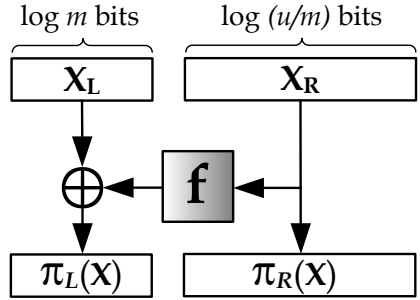

**Figure 3:** The one-round Feistel permutation used in our construction.

Then, in every bin we apply the scheme from Section 5.1 that relies on truly random permutations, but using three $k$-wise $\delta$-dependent permutations that are shared among all bins (recall that the latter scheme requires three permutations: one for its first-level hashing, and two for its permutation-based cuckoo hashing that stores the overflowing elements)[12]. By setting $k = n^{1/10} + n^{3/40}$ it holds that the distribution inside every bin is $\delta$-close in statistical distance to that when using truly random permutations. Therefore, Lemma 5.2 and Claim 5.3 guarantee that these permutations provide the required performance guarantees for each bin with probability $1 - (2^{-\omega(\log n)} + \delta) = 1 - 1/\text{poly}(n)$. Thus, applying a union bound on all $m$ bins implies that our construction provides the same performance guarantees as the one in Section 5.1 with probability $1 - 1/\text{poly}(n)$, for an arbitrary large polynomial.

We note that a possible (but not essential) refinement is to combine the queues of all $m$ bins. Recall that each bin has two queues: a queue for its de-amortized cuckoo hashing (see Section 3), and a queue for its first-level bins (see Section 4). An analysis almost identical to that of [ANS09] (for the de-amortized cuckoo hashing) and of Section 4 (for the first-level bins) shows that we can in fact combine all the queues of the de-amortized cuckoo hashing schemes, and all the queues of the first-level bins.

We are only left to prove that with high probability no bin contains more than $d = n^{1/10} + n^{3/40}$ elements:

**Claim 5.4.** *Fix $u$ and $n \leq u$, let $m = n^{9/10}$, and let $\mathcal{F}$ be a collection of $k'$-wise independent functions $f : \{0,1\}^{\log(u/m)} \to \{0,1\}^{\log m}$ for $k' = \lfloor n^{1/20}/\epsilon^{1/3} \rfloor$. For any set $S \subset \{0,1\}^{\log u}$ of size $n$, with probability $1 - 2^{-\omega(\log n)}$ over the choice of $f \in \mathcal{F}$, when using the function $x \mapsto x_L \oplus f(x_R)$ for mapping the elements of $S$ into $m$ bins, no bin contains more than $n^{1/10} + n^{3/40}$ elements.*

**Proof.** Without loss of generality we bound the number of elements that are mapped into the first bin, and the claim then follows by applying the union bound over all $m$ bins. Given a set $S$ of size $n$ we partition it into disjoint subsets $S_1, \ldots, S_t$ according to the $x_R$ values of its elements (i.e., two elements belong to the same $S_i$ if and only if they share the same $x_R$ value). Any element $x$ is mapped into the bin $x_L \oplus f(x_R)$, and therefore from each subset $S_i$ at most one element can be mapped into the first bin. For every $i \in [t]$ denote by $Y_i$ the indicator of the event in which an element of $S_i$ is mapped into the first bin, and let $Y = \sum_{i \in [t]} Y_i$. Then for every $i \in [t]$ it holds that $E(Y_i) = |S_i|/m$, and thus $E(Y) = n/m = n^{1/10}$. Since each subset $S_i$ corresponds to a different $x_R$ value, we have that the $Y_i$'s are $k'$-wise independent. Therefore, we can apply a Chernoff bound for random variables with limited independence due to Schmidt et al. [SSS95, Theorem 5.I.a]. Their bound guarantees that for independence $k' = \lfloor (n^{-1/40})^2 \cdot n^{1/10}/\epsilon^{1/3} \rfloor$, the probability that $Y > n^{1/10} + n^{3/40}$ is at most $e^{-\lfloor k'/2 \rfloor} = e^{-\Omega(n^{1/20})}$. ∎

Finally, we compute the total space consumption of the construction. The representations of the $k'$-wise independent function and the three $k$-wise $\delta$-dependent permutations require only a negligible number of bits compared to $\epsilon \mathcal{B}$. In addition, the scheme uses $m = n^{9/10}$ bins, each containing at most $d = n^{1/10} + n^{3/40}$ elements that are represented with $(1 + \epsilon) \log \binom{u/m}{d}$ bits using

---

[12] When dealing with a universe of size $u \leq n^{1+\gamma}$ for a small constant $\gamma < 1$, we can even store three *truly random* permutations, but this solution does not extend to the more general case where $u/m$ might be rather large.

the scheme from Section 5.1. Therefore, their space consumption is at most

$$m \cdot (1+\epsilon) \log \binom{u/m}{d} = (1+\epsilon) \log \left( \binom{u/m}{d}^m \right)$$

$$\leq (1+\epsilon) \log \binom{u}{md} \qquad (5.4)$$

$$= (1+\epsilon) \log \binom{u}{(1+n^{-1/40})n}$$

$$\leq (1+\epsilon) \log \left( \binom{u}{n} \left(\frac{u}{n}\right)^{n^{-1/40}n} \right) \qquad (5.5)$$

$$\leq (1+\epsilon) \log \binom{u}{n} + 2n^{-1/40} \log \left(\frac{u}{n}\right)^n$$

$$\leq (1+\epsilon) \log \binom{u}{n} + 2n^{-1/40} \log \binom{u}{n}$$

$$\leq (1+2\epsilon)\mathcal{B} \;,$$

where Equation (5.4) follows from the inequality $\binom{n_1}{k_1}\binom{n_2}{k_2} \leq \binom{n_1+n_2}{k_1+k_2}$, and Equation (5.5) follows from the fact that

$$\frac{\binom{u}{(1+\epsilon)n}}{\binom{u}{n}} = \frac{(u-n)\cdots(u-(1+\epsilon)n+1)}{((1+\epsilon)n)\cdots(n+1)} \leq \left(\frac{u}{n}\right)^{\epsilon n} \;.$$

We now explain how to deal with the more general case in which $u$ and $m$ are not powers of 2. In fact, $m$ can be chosen as the smallest power of two that is larger than $n^{9/10}$ (and $d$ is then adapted accordingly), and therefore we only need to handle $u$. Dealing with the one-round Feistel permutation is similar to Section 5.1.1. If $m$ divides $u$ then the construction extends to $\pi(x) = (x_L + f(x_R) \mod m, x_R)$, where $x_L = \lfloor \frac{x}{u/m} \rfloor$ and $x_R = x \mod u/m$. If $m$ does not divide $u$, then we decrease the size of the universe to $u' > u - m$ by ignoring at most $m - 1$ elements, such that $m$ divides $u'$. We store these ignored elements (if ever encountered) using a separate de-amortized cuckoo hashing scheme. There are less than $m < 2n^{9/10}$ such elements, and therefore the additional space consumption is only $O(n^{9/10} \log u) = O(n^{9/10} \log n)$ (recall that we can always assume that $u \leq n^c$, for a sufficiently large constant $c > 1$, by hashing the universe into a set of size $n^c$ using a pairwise independent hash function).

Dealing with the bins depends on the ratio between $u$ and $n$ (recall that inside the bins the elements are taken from a universe of size $u/m$). If $u \leq n^{1+\gamma}$ for some small constant $\gamma < 1$, then in fact we can afford to explicitly store three truly random permutations over a universe of size $u/m$ and continue exactly as in Section 5.1.1. In addition, if $u > n^{1+\alpha}$ then the space consumption in each of the bins is in fact $(1 + o(1))m \log(u/m)$ (see Case 2 in Section 5.1.4), and therefore we can allow ourselves to increase the size of the universe to $u'$ which is the smallest power of 2 that is larger than $u$ ($u' < 2u$ and this hardly affects the space consumption). Now both $u'$ and $m$ are powers of two, and therefore $u'/m$ is a power of 2, which means that we can use the $k$-wise $\delta$-dependent permutations described in Section 5.3.

## 5.3 $k$-Wise $\delta$-Dependent Permutations with Short Descriptions and Constant Evaluation Time

There are several known constructions of $k$-wise $\delta$-dependent functions with short descriptions and constant evaluation time (see Section 2). Naor and Reingold [NR99, Corollary 8.1], refining the framework of Luby and Rackoff [LR88], showed how to construct $k$-wise $\delta'$-dependent permutations from $k$-wise $\delta$-dependent functions. In terms of description length, each permutation in their collection consists of two pairwise independent permutations, and two $k$-wise $\delta$-dependent functions. Similarly, in terms of evaluation time, their construction requires two evaluations of pairwise independent permutations and two evaluations of $k$-wise $\delta$-dependent functions. Thus, by combining these results we obtain the following corollary:

**Corollary 5.5** ([NR99, Sie04]). *For any $n$, $w = O(\log n)$, and constant $c > 1$, there exists a polynomial-time algorithm outputting a collection $\Pi$ of permutations over $\{0,1\}^w$ with the following guarantees:*

1. *With probability $1 - n^{-c}$ the collection $\Pi$ is $k$-wise $\delta$-dependent, where $k = n^\alpha$ for some constant $\alpha < 1$ (that depends on $n$ and $w$), and $\delta = \frac{k^2}{2^{w/2}} + \frac{k^2}{2^w}$.*

2. *Any permutation $\pi \in \Pi$ can be represented using $n^\beta$ bits, for some constant $\alpha < \beta < 1$, and evaluated in constant time in the unit cost RAM model.*

As discussed in Section 2, the restriction to a polynomial-size domain does not hurt the generality of our results: in our applications the domain can always be assumed to be of sufficiently large polynomial size by using a pairwise (almost) independent function mapping it to a set of polynomial size without any collisions with high probability. In addition, for our schemes we need to use $\delta = 1/\text{poly}(n)$, which might be significantly smaller than $k^2/2^{w/2}$. Kaplan, Naor, and Reingold showed that composing $t$ permutations that are sampled from a collection of $k$-wise $\delta$-dependent permutations results in a collection of $k$-wise $(O(\delta))^t$-dependent permutations. Specifically, given a collection $\Pi$ of permutations and an integer $t$, let $\Pi^t = \{\pi_1 \circ \cdots \circ \pi_t\}_{\pi_1,\ldots,\pi_t \in \Pi}$, then:

**Theorem 5.6** ([KNR09]). *Let $\Pi$ be a collection of $k$-wise $\delta$-dependent permutations, then for any integer $t$, $\Pi^t$ is a collection of $k$-wise $(\frac{1}{2}(2\delta)^t)$-dependent permutations.*

Finally, we note that the above corollary shows that we can deal with roughly any $k < 2^{w/4}$ (in addition to the restriction $k = n^\alpha$ where the constant $\alpha$ provided by Siegel's construction). More generally, for any constant $0 < \gamma < 1/2$, Naor and Reingold presented a variant of their constructions that allows $k < 2^{w(1/2-\gamma)}$ that consists of essentially $1/\gamma$ invocations of the $k$-wise independent functions. This generalization, however, is not required for our application.

### 5.4 Using More Efficient Hash Functions

As discussed in Section 2, whenever we use $k$-wise independent functions in our construction (for $k = n^\alpha$ for some constant $0 < \alpha < 1$), we instantiate them with Siegel's construction [Sie04] and it simplification due to Dietzfelbinger and Rink [DR09]. The approach underlying Siegel's construction is currently rather theoretical, and for the case of full independence simpler and more efficient constructions were proposed by Dietzfelbinger and Woelfel [DW03] following Pagh and Pagh [PP08]. These constructions, however, provide a weaker guarantee than $k$-wise independence: For any *specific* set $S$ of size $k$, there is an arbitrary polynomially small probability of failure (i.e., choosing a "bad" function for this set), but if failure does not occur, then a randomly chosen function is fully random on $S$.

In what follows we prove that our scheme can in fact rely on such a weaker guarantee, resulting in significantly more efficient instantiations. Specifically, we show that Theorem 5.1 holds even when instantiating our scheme with the functions of Dietzfelbinger and Woelfel [DW03] (i.e., the scheme is exactly the same one, except for the hash functions). There are two applications of $k$-wise independent functions in our scheme. The first is the one-round Feistel permutation used for mapping the elements into first-level bins of size roughly $n^\alpha$ each. The second is the construction of $k$-wise $\delta$-dependent permutations which are used for handling the elements insides the first-level bins. We deal with each of these applications separately.

**Application 1: one-round Feistel permutation.** In this case we need to prove that Claim 5.4 holds even when the collection $\mathcal{F}$ of functions satisfies the weaker randomness guarantee discussed above. The main difference is that now the error probability will be $1/\text{poly}(n)$ for any pre-specified polynomial, instead of $2^{-\omega(\log n)}$ as in Claim 5.4. Note that we can allow ourselves to even use $k = n/\log^2 n$ in the construction of [DW03], since such functions will be described using $O(n/\log n)$ bits (see Section 2) which do not hurt our memory consumption. We prove the following claim:

**Claim 5.7.** *Fix any integers $u$ and $n \leq u$, let $m = n^{9/10}$, and let $\mathcal{F}$ be a collection of functions $f: \{0,1\}^{\log(u/m)} \to \{0,1\}^{\log m}$ with the following property: for any set $S' \subset \{0,1\}^{\log u}$ of size $k = n/\log^2 n$ it holds that with probability $1 - n^{-c}$ the values of a randomly chosen $f \in \mathcal{F}$ are uniform on $S'$. Then, for any set $S \subset \{0,1\}^{\log u}$ of size $n$, with probability $1 - n^{-(c-1)}$ over the choice of $f \in \mathcal{F}$, when using the function $x \mapsto x_L \oplus f(x_R)$ for mapping the elements of $S$ into $m$ bins, no bin contains more than $n^{1/10} + n^{1/20} \log n$ elements.*

*Proof.* Given a set $S$ of size $n$, we partition it arbitrarily to $n/k = \log^2 n$ subsets $S_1, \ldots, S_{n/k}$ of size $k = n/\log^2 n$ each (for simplicity we assume that $k$ divides $n$, but this is not essential for our proof). Then, with probability at least $1 - \frac{\log^2 n}{n^c}$, it holds that a randomly chosen $f \in \mathcal{F}$ is uniform on each of these subsets (note, however, that the values on different subsets are not necessarily independent). In this case, the same analysis as in Claim 5.4 (this time using a Chernoff bound for full independence), shows that with probability $1 - 2^{-\omega(\log n)}$, no bin contains more than $\frac{k}{n^{9/10}} + \left(\frac{k}{n^{9/10}}\right)^{1/2}$ elements from each subset. Therefore, summing over all $n/k$ subsets, the number of elements mapped to each bin is at most

$$\frac{n}{k} \cdot \left( \frac{k}{n^{9/10}} + \left(\frac{k}{n^{9/10}}\right)^{1/2} \right) = n^{1/10} + n^{1/20} \log n \ .$$

∎

**Application 2: first-level bins.** This case is much simpler since all we need is a function that behaves almost randomly on the specific sets of elements that are mapped to each bin (recall that each bin contains roughly $n^\alpha$ elements). Therefore, the type of guarantee provided by [DW03] together with a union bound over all bins are clearly sufficient. We obtain the following corollary as an alternative to Corollary 5.5:

**Corollary 5.8** ([NR99, DW03]). *For any $n$, $k \leq n$, $w = O(\log n)$, and constant $c > 1$, there exists a polynomial-time algorithm outputting a collection $\Pi$ of permutations over $\{0,1\}^w$ with the following guarantees:*

1. *For any set $S \subset \{0,1\}^w$ of size $k$, with probability $1 - n^{-c}$ for a randomly chosen permutation $\pi \in \Pi$, the distribution of the values of $\pi$ on $S$ is $\delta$-close to the distribution of the values of a truly random permutation on $S$, where $\delta = \frac{k^2}{2^{w/2}} + \frac{k^2}{2^w}$.*

2. *Any permutation $\pi \in \Pi$ can be represented using $O(k \log n)$ bits, and evaluated in constant time in the unit cost RAM model.*

On one hand the above corollary is weaker than Corollary 5.5 in terms of the guarantee on the randomness, as discussed in Section 2. On the other hand, however, when setting $k = n^\alpha$ the number of bits required to describe a function is only $O(n^\alpha \log n)$ compared to $n^\beta$ for some constant $\alpha < \beta < 1$ in Corollary 5.5. In turn, this allows to use slightly larger first-level bins which yields a better space consumption. In addition, the construction stated in the above corollary enjoys the same advantages of [DW03] over [Sie04], and in particular a better evaluation time.

We note that as in Section 5.3, Theorem 5.6 can be applied to reduce the value of $\delta$ in the above corollary to any polynomially small desirable value, by composing a constant number of such permutations (as long as $k < 2^{w(1/4-\gamma)}$ for some constant $0 < \gamma < 1/4$).

# 6 Concluding Remarks and Open Problems

**Implications of our constructions for the amortized setting.** We note that our constructions offer various advantages over previous constructions even in the amortized setting, where one is not interested in worst-case guarantees. In particular, instantiating our dictionary with the classical cuckoo hashing [PR04] (instead of its de-amortized variant [ANS09]) already gives a logarithmic upper bound *with high probability* for the update time, together with a space consumption of $(1 + \epsilon)n$ memory words for a *sub-constant* $\epsilon$.

**On the practicality of our schemes.** In this paper we concentrated on showing that it is possible to obtain a succinct representation with worst-case operations. The natural question is how applicable these methods are. There are a number of approaches that can be applied to reduce the overflow of the first-level bins. First, we can use the two-choice paradigm (or, more generally, $d$-choice) in the first-level bins instead of the single function we currently employ. Another alternative is to apply the generalized cuckoo hashing [DW07] inside the first-level bins, limiting the number of moves to a small constant, and storing the overflowing elements in de-amortized cuckoo hashing as in our actual construction. Experiments we performed indicate that these approaches result in (sometimes quite dramatic) improvements. The experiments suggest that for the latter variant, maintaining a small queue of at most logarithmic size, enables us even to get rid of the second-level cuckoo hashing: i.e., an element can reside in one of two possible first-level bins, or in the queue.

Another natural tweak is using a single queue for all the de-amortizations together. Finally, while the use of chopped permutations introduces only a negligible overhead, the use of an intermediate level seems redundant and we conjecture that better analysis would indeed show that.

**Clocked adversaries.** The worst-case guarantees of our dictionary are important if one wishes to protect against "clocked adversaries", as mentioned in Section 1. This in itself can yield a solution in the following sense: have an upper bound $\alpha$ on the time each memory access takes, and then make sure that all requests are answered in time exactly $\alpha$ times the worst-case upper bound on the number of memory probes. Such an approach, however, may be quite wasteful in terms of computing resources, since we are not taking advantage of the fact that some operations may be processed in time that is below the worst-case guarantee. In addition, this approach ignores the memory hierarchy, that can possibly be used to our advantage.

**Lower bounds for dynamic dictionaries.** The worst-case performance guarantees of our constructions are satisfied with all but an arbitrary small polynomial probability over the randomness of their initialization phase. There are several open problems that arise in this context. One problem is to reduce the failure probability to sub-polynomial. The main bottleneck is the approximation to $k$-wise functions or permutations. Another bottleneck is the lookup procedure of the queue (if the universe is of polynomial size then we can in fact maintain a small queue deterministically). Another problem is to identify whether randomness is needed at all. That is, whether it is possible to construct a deterministic dictionary with similar guarantees. We conjecture that randomness is necessary. Various non-constant lower bounds on the performance of deterministic dynamic dictionaries are known for several models of computation [DKM+94, MNR90, Sun91]. Although these models capture a wide range of possible constructions, for the most general cell probe model [Yao81] it is still an open problem whether a non-constant lower bound can be proved[13].

**Extending the scheme to smaller values of $\epsilon$.** Recall that in the de-amortized construction of perfect hashing inside the first-level bins (Section 4), we suggested a specific scheme that can handle $\epsilon = \Omega((\log \log n / \log n)^{1/2})$. This restriction on $\epsilon$ was dictated by the space consumption of the global lookup tables together with the hash functions inside each bin. The question is how small can $\epsilon$ be and how close to the information theoretic bound can we be, that is for what function $f$ can we use $\mathcal{B} + f(n, u)$ bits. A possible approach is to use the two-choice paradigm for reducing the number of overflowing elements from the first level of our construction, as already mentioned.

**Constructions of $k$-wise almost independent permutations.** In Section 5.3 we observed a construction of $k$-wise $\delta$-dependent permutations with a succinct representation and a constant evaluation time. Two natural open problems are to allow larger values of $k$ (the main bottlenecks are the restrictions $k < u^{1/2}$ in [NR99] and $k \leq n^\alpha$ in [Sie04]), and a sub-polynomial $\delta$ (the main bottleneck is the failure probability of Siegel's construction [Sie04]).

**Supporting dynamic resizing.** In this paper we assumed that there is a pre-determined bound on the maximal number of stored elements. It would be interesting to construct a dynamic dictionary with constant worst-case operations and full memory utilization *at any point in time*. That is, at any point in time if there are $\ell$ stored elements then the dictionary occupies $(1 + o(1))\ell$ memory words (even more challenging requirement may be to use only $(1 + o(1))\mathcal{B}(u, \ell)$ bits of memory, where $\mathcal{B}(u, \ell)$ is the information-theoretic bound for representing a set of size $\ell$ taken from a universe of size $u$). This requires designing a method for dynamic resizing that essentially does not incur any noticeable time or space overhead in the worst case. We note that in our construction it is rather simple to dynamically resize the bins in the first-level table, and this provides some flexibility.

**Dealing with multisets.** A more general variant of the problem considered in this paper is constructing a dynamic dictionary that can store multisets of $n$ elements taken from a universe of size $u$. In this setting the information-theoretic lower bound is $\log \binom{u+n}{n}$ bits. Any such dictionary with a succinct representation and constant-time operations in the worst case can be used to construct a Bloom filter alternative that can also support deletions (similar to Appendix A). This will improve the result obtained by the construction of Pagh et al. [PPR05] that supports deletions, but guarantees constant-time operations only on amortized, and not in the worst case.

---
[13]There is an unpublished manuscript of Rajamani Sundar from 1993 titled "A lower bound on the cell probe complexity of the dictionary problem", reported by Miltersen [Mil99] and Pagh [Pag02]. To the best of our knowledge, the nature of this result is currently unclear.

## Acknowledgments
We thank Rasmus Pagh and Udi Wieder for many useful remarks and suggestions.

[Woe06]     P. Woelfel. Maintaining external memory efficient hash tables. In *10th International Workshop on Randomization and Computation*, pages 508–519, 2006.

[Yao81]     A. C.-C. Yao. Should tables be sorted? *Journal of the ACM*, 28(3):615–628, 1981.

[ZLP08]     B. Zhu, K. Li, and R. H. Patterson. Avoiding the disk bottleneck in the data domain deduplication file system. In *Proceedings of the 6th USENIX Conference on File and Storage Technologies*, pages 269–282, 2008.

## A  Application of Small Universes: A Nearly-Optimal Bloom Filter Alternative

In this section we demonstrate an application of our succinctly-represented dictionary that uses a rather small universe, for which the difference between using $(1 + o(1))\log \binom{u}{n}$ bits and using $(1 + o(1))n \log u$ bits is significant. We consider the *dynamic approximate set membership problem*: representing a set $S$ of size $n$ defined dynamically via a sequence of insertions, in order to support lookup queries, allowing a false positive rate of at most $0 < \delta < 1$, and no false negatives. That is, the result of a lookup query for any element $x \notin S$ is correct with probability at least $1 - \delta$, and the result of a lookup query for any element $x \in S$ is always correct. In both cases the probability is taken only over the randomness of the data structure. The information-theoretic lower bound for the space required by any solution to this problem is $n \log(1/\delta)$ bits, and this holds even in the static setting where the set is given in advance [CFG+78].

This problem was first solved using a Bloom filter [Blo70], a widely-used data structure proposed by Bloom (see the survey by Broder and Mitzenmacher [BM03] for applications of Bloom filters). Bloom filters, however, suffer from various weaknesses, mostly notably are the dependency on $\delta$ in the lookup time which is $\log(1/\delta)$, and the sub-optimal space consumption which is $n \log(1/\delta) \log e$ bits. Over the years extensive research was devoted for improving the performance of Bloom filters in the static case (e.g., [Mit02, DP08, Por09]), as well as for the closely related *retrieval problem* (e.g., [CKR+04, DMadHP+06, DP08]).

**A general solution using a dictionary.**  Carter et al. [CFG+78] proposed a general method for solving the above problem using any dictionary: given a set $S = \{x_1, \ldots, x_n\} \subseteq \mathcal{U}$, sample a function $h : \mathcal{U} \to [n/\delta]$ from a collection $\mathcal{H}$ of universal hash functions, and use the dictionary for storing the set $h(S) = \{h(x_1), \ldots, h(x_n)\}$. The correctness of the dictionary guarantees that the result of a lookup query for any element $x \in S$ is always correct. In addition, for any element $x \notin S$ is holds that

$$\Pr\left[h(x) \in h(S)\right] \leq \sum_{i=1}^{n} \Pr\left[h(x) = h(x_i)\right] = n \cdot \frac{\delta}{n} = \delta \ .$$

Therefore, the result of a lookup query for any element $x \notin S$ is correct with probability at least $1 - \delta$ over the choice of $h \in \mathcal{H}$. Note that in case that the dictionary supports insertions (as in our case), then the elements of the set $S$ can by provided one by one.

This approach was used by Pagh et al. [PPR05] who constructed an alternative to Bloom filters by relying on the dictionary of Raman and Rao [RR03]. Their construction uses $(1 + o(1))n \log(1/\delta) + O(n + \log u)$ bits of storage, guarantees constant lookup time which is independent of $\delta$, and supports insertions and deletions in amortized expected constant time (that is, they actually solve a more general variant of this problem that deals with multisets). Another feature of their construction, which is also shared by our construction, is the usage of explicit hash functions, as opposed to assuming the availability of a truly random hash function as required for the analysis of Bloom filters.

**Using our succinctly-represented dictionary.** Using our succinctly-represented dictionary and the method of Carter et al. [CFG+78] we immediately obtain an alternative to Bloom filters, which uses $(1 + o(1))n \log(1/\delta) + O(n + \log u)$ bits[14], guarantees constant lookup time which is independent of $\delta$, and supports insertions in constant time (independent of $\delta$) in the worst case with high probability. As pointed out in Section 1.1, for any sub-constant $\delta$, and under the reasonable assumption that $u \leq 2^{O(n)}$, the space consumption is $(1 + o(1))n \log(1/\delta)$, which is optimal up to an additive lower order term.

## B  Negatively Related Random Variables

In the proofs of Lemmata 3.2 and 5.2 we apply Chernoff bounds on the sum of indicator random variables that are not independent, but are *negatively related* as defined by Janson [Jan93], who showed that these bounds are indeed applicable in such a setting.

**Definition B.1** ([Jan93])**.** Indicator random variables $(I_i)_{i=1}^m$ are *negatively related* if for every $j \in [m]$ there exist indicator random variables $(J_{i,j})_{i=1}^m$ defined on the same probability space (or an extension of it), such that:

1. The distribution of $(J_{i,j})_{i=1}^m$ is identical to that of $(I_i)_{i=1}^m$ conditioned on $I_j = 1$.

2. For every $i \neq j$ it holds that $J_{i,j} \leq I_i$.

In the proof of Lemma 3.2 we consider an experiment in which $n$ balls are mapped independently and uniformly at random into $m$ bins. For every $i \in [m]$ denote by $I_i$ the indicator of the event in which the $i$-th bin contains at least $t$ balls, for some threshold $t$ (dealing with the case of bins with at most $t$ balls is essentially identical). We now argue that the indicators $(I_i)_{i=1}^m$ are negatively related by defining the required indicators $(J_{i,j})_{i=1}^m$ for every $j \in [m]$. Consider the following experiment: map $n$ balls into $m$ bins independently and uniformly at random, and define $(I_i)_{i=1}^m$ accordingly. If the $j$-th bin contains at least $t$ balls then define $J_{i,j} = I_i$ for every $i \in [m]$. Otherwise, denote by $T$ the number of balls in the $j$-th bin, and sample an integer $T'$ from the distribution of the number of balls in the $j$-th bin conditioned on having at least $t$ balls in that bin. Choose uniformly at random $T' - T$ balls from the balls outside the $j$-th bin, and move them to the $j$-th bin. Define $(J_{i,j})_{i=1}^m$ according to the current allocation of balls into bins (i.e., $J_{i,j} = 1$ if and only if the $i$-th bin contains at least $t$ balls). Then, the independence between different balls implies that the indicators $(J_{i,j})_{i=1}^m$ have the right distribution, and that for every $i \neq j$ it holds that $J_{i,j} \leq I_i$ since we only removed balls from other bins.

In the proof of Lemma 5.2 we consider a similar experiment where the mapping of balls into bins is done using a chopped permutation $\pi$ over $\mathcal{U}$. The above argument extends to this setting, with the only difference that moving balls from one bin to another is a bit more subtle. Specifically, for moving $T' - T$ balls to the $j$-th bin, we first randomly choose $T' - T$ values $y_1, \ldots, y_{T'-T} \in \mathcal{U}$ that belong to the $j$-th bin (after the chopping operation) and their $\pi^{-1}$ values are not among the $n$ balls (these $y_i$'s correspond to empty locations in the $j$-th bin). Then, we randomly choose $x_1, \ldots, x_{T'-T} \in \mathcal{U}$ from the set of balls that were mapped into other bins, and for every $1 \leq i \leq T' - T$ we switch between the values of $\pi$ on $x_i$ and $\pi^{-1}(y_i)$.

---

[14]Specifically, the dictionary uses $(1 + o(1))n \log \binom{n/\delta}{n} \leq (1 + o(1))n \log(1/\delta) + O(n)$ bits, and the universal hash function is described using $2\lceil \log u \rceil$ bits.